\documentclass[preprint]{aastex}

\bibpunct[]{(}{)}{;}{a}{}{;}

\newcommand{\sub}[1]{_{\mbox{{\scriptsize #1}}}}
\newcommand \simg{\hspace{1ex} ^{>} \hspace{-2.5mm}_{\sim} \hspace{1ex}}
\newcommand \siml{\hspace{1ex} ^{<} \hspace{-2.5mm}_{\sim} \hspace{1ex}}
\begin{document}

\title{The Effects of a Stellar Encounter on a Planetesimal Disk}
\author{Hiroshi Kobayashi and Shigeru Ida}
\affil{Department of Earth and Planetary Sciences,\\
Tokyo Institute of Technology}
\affil{Meguro-ku, Tokyo 152-8551, Japan}
\email{hkobayas@geo.titech.ac.jp}

\begin{abstract}
We investigate the effects of a passing stellar encounter 
 on a planetesimal disk through analytical calculations and numerical
 simulations, and derive the boundary radius 
 ($a_{\rm planet}$) outside which planet formation is inhibited
 by disruptive collisions with high relative velocities. 
Ida, Larwood, and Burkert (2000. ApJ. 528, 1013-1025) 
suggested that a stellar encounter caused inhibition of planet formation 
in the outer part of a protoplanetary disk. 
We study orbital eccentricity ($e$) and
 inclination ($i$) of planetesimals pumped up
 by perturbations of a passing single star.
We also study the degree of alignment of longitude of pericenter  
and ascending node to estimate relative velocities between the planetesimals. 
We model a protoplanetary system as a disk of massless particles
 circularly orbiting a host star, following \citet{ida00}.
 The massless particles represent planetesimals.
A single star as massive as the host star encounters the
 protoplanetary system.
Numerical orbital simulations show that
 in the inner region at semimajor axis $a \la 0.2 D$ where $D$
 is pericenter distance of the encounter, 
$e$ and $i$ have power-law dependence on $(a/D)$ as 
 $e \propto (a/D)^{5/2}$ and $i \propto (a/D)^{3/2}$ 
and the longitudes are aligned, 
independent of the encounter parameters.
In the outer region $a \ga 0.2 D$, the radial gradient is steeper,
 and is not expressed by a single power-law.
The longitudes are not aligned. 
Since planet accretion is inhibited by $e$ as small as 0.01, 
we focus on the weakly perturbed 
 inner region. 
We analytically reproduce the power-law dependence and explicitly give
 numerical factors of the power-law dependence as functions of
 encounter parameters.
We derive the boundary radius 
 ($a_{\rm planet}$) of planet forming region as a function of
 dynamical parameters of a stellar cluster, 
assuming the protoplanetary system belongs to the stellar cluster. 
Since the radial gradient of $e$ is so steep that the boundary is
 sharply determined.
Planetesimal orbits are significantly modified beyond the boundary,
 while they are almost intact inside the boundary. 
This tendency is strengthened by reduction of relative velocity 
due to the longitude alignment in the inner region. 
We find $a_{\rm planet} \sim 40$-60AU in the case of $D \sim 150$-200AU.
$D \sim $ 200AU may be likely to occur in a
 relatively dense cluster.
We point out that the size of planetary systems ($a_{\rm planet}$) 
born in a dense cluster may
 be necessarily restricted to that comparable to the size of
 planet region ($\sim 30$-40AU) of our Solar system.
\end{abstract}

\newpage
\newpage
\section{INTRODUCTION}

In general, stars are born as members of an open cluster.
 Stellar clusters would evaporate on timescales 
 more than $10^8$ years \citep{kroupa1995,kroupa1998}.
This evaporation would be caused by gravitational interactions
 between stars, so that many stars experience gravitational 
 perturbations of the other stars during the evaporation.
More than half of T Tauri stars have protoplanetary disks
 \citep[e.g.,][]{beckwith1996}, which would eventually form planetary
 systems on timescales $10^{6}$-$10^{9}$ years 
\citep*[e.g.,][]{safronov69, wetherill80,hayashi85}.
Planetary systems would be affected by stellar
 encounters more or less during their formation stage.

In the standard model 
\citep[e.g.,][]{safronov69, wetherill80,hayashi85,lissauer93},
terrestrial planets and
 cores of jovian planets accrete from planetesimals that 
 are formed in a protoplanetary disk.
The accretion of jovian planet cores is followed by
 gas accretion onto the core when the core acquires
 critical mass $\sim$ 5-15 $M_\oplus$ 
\citep[e.g.,][]{mizuno80,bodenheimer86,ikoma00}. 

The passing stellar encounters would pump up orbital eccentricity $e$ 
 and inclination $i$ of planetesimals.  
The velocity dispersion of planetesimals is given by
 $\sim \sqrt{e^2 + i^2} v_{\rm kep}$, where $i$ is given in unit of radian
 and $v_{\rm kep}$ is Keplerian velocity 
\citep[e.g.,][]{safronov69, lissauer93, ohtsuki93}.  
If the velocity dispersion exceeds their surface escape velocity of
 planetesimals, 
a collision between the planetesimals results in disruption rather than accretion 
 because of the high velocity collision unless their pericenters are aligned 
\citep[e.g.,][; also see section 2]{safronov69,greenberg78,ohtsu93}.
Then, planetesimal accretion would be forestalled.
As shown in Eq. (\ref{eq:e_bound}) in section 2, 
if pumped-up $e$ and $i$ are larger than 0.01, 
the velocity dispersion exceeds the surface escape velocity of
 planetesimals in the early stage.
Therefore, small orbital modification ($e,i \ga 0.01$)
 can give significant influence on planet formation. 
If the longitudes of pericenter and ascending nodes of colliding planetesimals 
are aligned (`` phase alignment''), the colliding velocity is significantly 
reduced from that in the above argument 
\citep[][; also see section 6]{marzari00}. Hence, we also examine the 
degree of the alignment. 

This work is motivated by Ida, Larwood, and Burkert (2000; hereafter ILB00). 
 They showed that the radial gradient of the pumped-up $e$ and $i$ is rather steep;
 $e$ and $i$ are highly pumped up in the outer planetesimal disk, while
 the inner disk is almost intact
(also see sections 3 and 4). 
The steep gradient leads to a sharp boundary of the disk that
 divides the strongly perturbed region where planet formation is
 inhibited and the intact region where planet formation keeps going.  
The boundary radius determines radial size of the region of a 
planetary system where planetary-sized bodies exist. 
Most planetesimals in the outermost region are ejected (and some of them
 are captured by the passing star), which would truncate the region of the 
planetary
 system where solid materials exist. 
Here we are concerned with the former size. 

In the present paper, we investigate the orbital modification 
 due to a passing stellar encounter and discuss the effects on
 planetary formation.
We derive analytical formulae of pumped up $e$ and $i$,
 because, as stated above, the boundary between the planet forming region
 and the planet-formation inhibited region is marked 
 by $e$, $i$  as small as 0.01 and hence linear perturbation analysis
 is available to determine the boundary.
We also did numerical simulations of 10,000 test particles in
 some parameter ranges.
Our analytical formulae show excellent agreement with the numerical
 results.

Encounters between a star and a particle/gas disk have been studied 
by many authors in different contexts.
However, most previous studies were concerned with the strongly
perturbed region in the disk.

Galactic encounters may form galactic tidal bridges and tails.
Many numerical simulations  \citep[e.g.,][and references therein]{tt72,barn92} 
have been done
 and analytical calculations with impulse approximation was applied
 for high speed galactic encounters \citep[e.g.,][]{binney1987}.
\citet{kalas95}
suggested that asymmetry in the dust disk 
around $\beta$ Pic is caused by a passing stellar encounter. 
\citet{kalas00} and \citet{john00} demonstrated through
 numerical simulation that a passing stellar encounter reproduces
 the observed disk asymmetry and ringlet structure.  
These structure in galactic and dust disks is 
formed in the outer part of a disk which is
 strongly perturbed.

A galaxy may be captured by another galaxy through 
energy transfer from orbital motions to internal motions caused by 
a close encounter 
to be a satellite galaxy \citep{plamer82, wahde96}.
Similarly, an encounter of a star with another star with 
 a protoplanetary gas disk may lead to capture of the passing star 
to form binary stars.  
It may also make a new companion from the disk \citep{boffin98},
 cause disk truncation \citep{hall97}, or cause rapid disk accretion
 \citep{ost94}.
The transfer of energy and angular momentum between a passing star 
 and the disk is calculated numerically
 \citep{clarke91, hall96, heller93, heller95}
 and analytically \citep{ost94}.
\citet{kory95} and \citet{larwood97} did more detailed analytical calculations.
These studies were mostly concerned with total energy and angular 
 momentum transfer, that is, concerned with 
strongly perturbed regions in the disk.

\citet{ost94} also studied weakly perturbed regime, 
although the author mainly focused on the strongly perturbed region where 
resonant effects are important. The author only presented approximately 
averaged 
change in angular momentum
 in the weakly perturbed region. 
As we discuss later, the author's results would not give enough information 
to deduce change in $e$ and $i$,
 which we want to use to discuss the effects on planet accretion.  

Passing stellar encounters are also important for planetary
 systems or planetesimal disks.
The planetesimals ejected by jovian planets 
 may become weakly bounded in the Solar system to form Oort cloud.
Passing stellar encounters would make their binding stronger
\citep{brunini96,fern97} or send them back to inner Solar system
 as long-period comets \citep{eggers96,yabu82}.
They considered stellar encounters with planetesimals 
 ejected from a protoplanetary system.

At present, mean distance between stars in the solar neighborhood
 is so large that only Oort cloud would be affected by passing stellar
 encounters.
However, since stars are generally born as members of an open cluster
 and stay in the cluster on timescales 
 more than $10^8$ years \citep{kroupa1995,kroupa1998},
 passing stellar encounters affect a planetary system or a 
protoplanetary planetesimal system itself in the early stage. 
Monte-Carlo numerical simulations suggest that
 the encounters can modify nearly circular orbits of giant planets 
to eccentric orbits,
 which may correspond to observed extrasolar planets in eccentric orbits
 \citep{marcos97,lau98,lau00,bonnell01}. 
ILB00 
 showed through numerical simulations 
that a stellar encounter may explain high orbital eccentricities and 
 inclinations of outer Kuiper belt objects. 
They suggested that planetary formation was inhibited by the high
 orbital eccentricities and inclinations in the Kuiper belt while
 more inner region was almost intact and planetary formation proceeded,
 if the most effective stellar encounter with the proto-Solar system
 had pericenter distance $\sim 150$-200AU. 
We perform more extensive calculations by both numerically and analytically 
to discuss the inhibition of planet formation in detail. 
\citet{heggie96} derived analytical formula for  
the change in eccentricity of binary stars by an 
encounter of a single passing star, 
using the Laplace-Runge-Lenz vector. 
In the limit of null mass of a binary companion, their formula is 
reduced to our analytical formula, although we used different method with 
the Gauss's equations \citep{brouwer}. 
We derive the change in not only eccentricity but also 
inclination and study relative velocity of planetesimals taking into account 
the alignment of longitudes. 
in order to apply to planetary formation. 

\citet{heppen}, \citet{whitmire} and \citet{marzari00}
study planet formation in a system with a 
binary companion orbiting outside a circumstellar disk around the primary star. 
The circumstellar disk is perturbed by the binary companion. 
They investigated relative velocity between planetesimals 
in the perturbed disk to find the condition of 
inhibition of planet formation, 
while we consider the effect of one passage of a field star. 

In section 2, we briefly summarize critical velocity dispersions 
or critical eccentricities 
of planetesimals that are important to discuss planet accretion. In section3, 
we explain calculation models.
In section 4, we show results of numerical simulations.
The analytical formulae are derived in section 5.
$e$ and $i$ pumped up by stellar perturbations in the weakly perturbed
 inner regime are
 explicitly given as functions of heliocentric radius and parameters
 of stellar encounters.  
Using the analytical formulae, in section 6,
 we discuss the size of planet forming
 region as a function of parameters of stellar clusters.
We will show the radius of planet forming region is likely to be
 40-60AU, in the case of a dense cluster, which may be consistent with
 the Solar system.

\section{PLANETARY ACCRETION}

Planetesimals with 1-10 km sizes are formed from dust grains 
in a plotoplanetary disk through self-gravitational instability 
\citep[e.g.,][]{safronov69, goldreich73} 
or sticking induced by turbulence of the disk gas 
\citep[e.g.,][]{weidenschilling93}. Accretion of the planetesimals make 
terrestrial planets and cores of jovian planets. 
The sticking (accretion) of planetesimals is caused by self-gravity. 
The accretion occurs, only when rebounding velocity of planetesimals 
after a collision is smaller than 
their surface escape velocity ($v\sub{esc}$). 
Since collision velocity is $\sim$ $\sqrt{v_0^2+v\sub{esc}^2}$ 
\citep{ohtsu93}, where $v_0$ is relative velocity of unperturbed crossing 
orbits
 neglecting mutual gravity, accretion occurs 
when $v_0 \la v\sub{esc}$, except when too elastic or too inelastic 
cases \citep{ohtsu93}. 
If stellar encounters pump up $v_0$ more than $v\sub{esc}$, 
a collision between planetesimals results in disruption rather than accretion 
\citep[e.g.,][]{safronov69,greenberg78}. 
Then planet formation is forestalled in this region. 
If the longitudes of pericenter and ascending node are randomly distributed, 
$v_{0}$ is nearly equal to velocity dispersion given by $v\sub{d}$ 
$\sim$ $\sqrt{e^2 + i~}$ $v\sub{kep}$ 
\citep[e.g.][]{safronov69, lissauer93, ohtsuki93}. 
As shown in section 6, the longitudes of the perturbed planetesimal orbits 
are not always random. 
We thus define ``effective'' eccentricity and inclination,
\begin{eqnarray}
e\sub{eff} &=& | v_{0,xy} |/v\sub{kep},\\
i\sub{eff} &=& | v_{0,z} |/v\sub{kep},
\end{eqnarray}
where $v_{0,xy}$ and $v_{0,z}$ are the component on the initial disk and 
the vertical to the disk, respectively.
By definition, $e\sub{eff} \la e$ and $i\sub{eff} \la i$. 
If the longitudes are aligned, $i\sub{eff} = 0$ and $e\sub{eff}$ 
expresses small relative velocity due to shear motion. 

The surface escape velocity of 
 planetesimals with mass $m$ and internal density $\rho$ is 
\begin{equation}
v\sub{esc} = \left( \frac{32 \pi G^3 m^2 \rho}{3} \right)^{\frac{1}{6}}.
\label{eq:def_v_esc}
\end{equation}
As shown in section 6, $e\sub{eff} \ga i\sub{eff}$, so that $v_0$ 
$\sim$  $e\sub{eff} v\sub{kep}$. 
The condition for a disruptive collision is 
$v_0 > v\sub{esc}$ $=$ $e\sub{eff,crit} v\sub{kep}$ 
\citep[e.g.,][]{starn}. 
The condition with effective eccentricity is 
\begin{equation}
e\sub{eff} > e\sub{eff,crit} \simeq
\left( \frac{32 \pi m^2 a^3 \rho}{3 M_1^3} \right)^{\frac{1}{6}},
\label{eq:e_bound_def}
\end{equation}
where $a$ and $M_1$ are semimajor axis of a planetesimal and 
the mass of the host star, respectively.
Numerically, $e\sub{eff,crit}$ is 
\begin{equation}
e\sub{eff,crit}
\sim 0.01 \left(\frac{m}{10^{22} \mbox{g}} \right)^{\frac{1}{3}} 
\left(\frac{\rho}{1 \mbox{g/cm}^3} \right)^{\frac{1}{6}} 
\left( \frac{a}{10 \mbox{AU}} \right)^{\frac{1}{2}}, 
\label{eq:e_bound}
\end{equation}
where the nominal value of $m \sim$ $10^{22}$ g corresponds to typical mass of 
Kuiper belt objects. 
In section 6, we show that the phase alignment breaks down for $e \ga 0.01$, 
so that $e\sub{eff} \simeq e$ for $e \ga 0.01$ and the condition 
(\ref{eq:e_bound_def}) is no other than the  condition for $e$. 
Equation (\ref{eq:e_bound}) shows that $e$ 
as small as 0.01 can significantly affect planetary accretion. 
If we consider ``protoplanets'' with lunar mass, $e\sub{eff,crit}$ 
is $\sim$ 0.1. 
However, it does not change the discussion on planet accretion in section 7.

\section{CALCULATION MODEL AND BASIC EQUATION}

We model a planetesimal disk as non-self-gravitating,
collisionless particles that initially have coplanar
circular orbits around a primary (host) star, 
because two-body relaxation time and mean collision time 
between planetesimals are much longer than an effective encounter time scale 
that is comparable to Kepler time scale at pericenter distance ($D$) 
of the encounter (for example, it is $\sim 10^3$ years for $D \sim 100$AU). 
We also neglect hydrodynamical gas drag, because 
the damping time due to the drag [$\simeq 10^7 ( m/10^{22} \mbox{g})^{1/3} 
(e/0.1)^{-1}$ yr at 10AU \citep{adachi}] is longer 
than the effective encounter time $\sim 10^3$ years for $m > 10^{10}$g at 10AU;
we are considering planetesimals with much larger masses. 
The particulate disk encounters a hypothetical passing star. 
Note that the gravitational relaxation, collision and the drag 
can be important when we consider planet formation on 
a longer time scale after the stellar passing. 
The equation of motion of a planetesimal in the heliocentric
frame (the frame with the primary star at center) is
\begin{equation}
\frac{d^2 {\bf r}_j}{dt^2} = - \frac{GM_1}{\mid {\bf r}_j \mid^3}
{\bf r}_j + \frac{GM_2}{\mid {\bf R}-{\bf r}_j \mid^3}
({\bf R}-{\bf r}_j) - \frac{G M_2}{\mid {\bf R} \mid^3}
{\bf R},
\label{eq:eq_motion}
\end{equation}
where $M_1$ and $M_2$ are masses of the primary and the passing stars, 
${\bf r}_j$ and ${\bf R}$ are position vectors of
the planetesimal $j$ and the passing star. 
The first term in the r. h. s. is force to produce Kepler motion around
the primary star, and the second and third terms are 
direct and indirect perturbing forces of the passing star.

We scale length by pericenter distance $D$ of the 
stellar encounter, mass by the primary
star mass $M_1$, and time by $\Omega\sub{kep}^{-1}$
where $\Omega\sub{kep}$ is Keplerian frequency at $a=D$ given by 
$\sqrt{G M_1/D^3}$.
Equation (\ref{eq:eq_motion}) is then transformed to
\begin{equation}
\frac{d^2 \tilde{\bf r}_j}{d\tilde{t}^2} = - \frac{\tilde{\bf r}_j}
{\mid \tilde{\bf r}_j \mid^3} - \frac{M_*(\tilde{\bf r}_j-\tilde{R})}
{\mid \tilde{\bf r}_j-\tilde{R} \mid^3} - \frac{M_*\tilde{R}}{\mid \tilde{R} \mid^3},
\label{eq:sc_eq_motion}
\end{equation}
where $M_* = M_2/M_1$, $\tilde{\bf r}_j ={\bf r }_j/D$,
$\tilde{\bf R} = {\bf R}/D$, and $\tilde{t} = \Omega\sub{kep} t$. 
Thus the parameters of encounters 
are inclination ($i_*$) relative to the initial planetesimal disk, 
eccentricity ($e_*$), and argument of perihelion ($\omega_*$) 
of orbit of the passing star, 
and the scaled
passing star mass ($M_*$).
The encounter geometry is illustrated in Fig. \ref{config}.
We calculate changes in $e$ and $i$ of the planetesimals
according to Eq. (\ref{eq:eq_motion}) or (\ref{eq:sc_eq_motion}) 
with various encounter parameters, 
through orbital integration 
and analytical estimation.

\section{NUMERICAL SIMULATION}

Regarding the method of numerical integration, 
we follow ILB00. 
We integrated orbits of 10,000 particles
with surface number density $n_{s} \propto a^{-3/2}$.
The particles are distributed in the region $a/D =$ 0.05-0.8.
Since we neglect mutual gravity and collisions of planetesimals, 
the particular choice of $a$-dependence of $n_s$ and  
outer and inner edges of the disk does not affect the results. 
The initial $e$ and $i$ of particles are 0. 
We integrated Eq. (\ref{eq:sc_eq_motion}), 
using a fourth order predictor-corrector scheme. 
Much more variations of encounter 
geometry, encounter velocity, and passing star mass were examined 
than ILB00 did. 

Figures \ref{time_d} show time evolution of
$e$ (left panels) and $i$ (middle panels) and 
corresponding face-on snapshots (right panels) 
in the
case with $e_* = 1$ (parabolic orbit), $i_* = 30^{\circ}$, 
$\omega_* = 0^{\circ}$, and $M_* = 1$.
The (a) top, (b) middle and (c) bottom panels show snapshots at 
$\tilde{t} =$ $- 1.33$, 0 and 1.33. 
The resultant $e$ and $i$ of planetesimals 
are mostly acquired when the passing star 
is near the pericenter.

The pumped-up $e$ and $i$ are shown in Fig. \ref{M_e_dep}, 
as a function of the scaled initial semimajor axis $a/D$, 
in the case with $i_* = 5^{\circ}$ and 
$\omega_* = 90^{\circ}$. ($M_*$, $e_*$) = (1,1), 
(0.2,1) and (1,5) in Figs. \ref{M_e_dep}a, 
\ref{M_e_dep}b and \ref{M_e_dep}c, respectively. 
In all cases, we find three characteristic regions of pumped-up $e$ and $i$. 
In the inner region at $a/D \la$ 0.1-0.3, 
$e$ and $i$ are in proportion to $(a/D)^{5/2}$ and $(a/D)^{3/2}$, 
respectively. 
In the outer region at $a/D$ $\ga$ 0.1-0.3, 
$e$ and $i$ have steeper $a$-gradient and divergence due to initial mean 
anomaly of planetesimals. 
In the outermost region, 
$e$ of many planetesimals, is greater than 1, that is,  
the particles are ejected from the system. 
ILB00 also found these features. 
We find that $e \propto (a/D)^{5/2}$ and $i \propto (a/D)^{3/2}$ 
always hold. Analytical estimate derived in section 5 and Appendix 
reproduces $e \propto (a/D)^{5/2}$ and $i \propto (a/D)^{3/2}$. 
Note that the dashed lines with triangles 
are analytical estimate derived in section 5 and Appendix. 
Figures 3 show larger $M_*$ and/or smaller $e_*$ produce 
larger $e$ and $i$. 
The agreement between the numerical and analytical results imply that 
$e$ and $i$ are scaled by $M_*/\sqrt{M_*+1}$. 
The dependence on $e_*$ is a more complicated form 
(see Appendix, Eqs. (\ref{eq:app_h_e}) to (\ref{eq:app_q_e})). 

Figures \ref{is_dep} show the dependence of $i_*$, 
in the case with
$\omega_* = 0^{\circ}$, $e_*=1$ and $M_*=1$.
$i_*$ are $5^{\circ}$ (Fig. \ref{is_dep}a), $30^{\circ}$ 
(Fig. \ref{is_dep}b), $45^{\circ}$ (Fig. \ref{is_dep}c), 
$85^{\circ}$ (Fig. \ref{is_dep}d) and $150^{\circ}$ (Fig. \ref{is_dep}e).
An encounter with  $ 0^{\circ} < i_* < 90^{\circ} $ is a prograde encounter 
relative to rotation of the planetesimal disk and 
one with $ 90^{\circ} < i_* < 180^{\circ}$ is retrograde. 
The numerical results show $i$ is dependent on $i_*$ like 
$\propto$ $\sin 2 i_*$ in the inner region. Comparison between 
Figs. \ref{is_dep}b and \ref{is_dep}f suggest that $e$ and $i$ are the same 
between $i_*$ and $90^{\circ}-i_*$ in the inner region, although 
retrograde encounters lead to less steep $a/D$-gradient than that in 
prograde encounters in the outer region ($a/D \ga$ 0.1-0.3). 
This shows that a secular effect is at work, so that 
we take the orbital average of the particle \citep{larwood97} in section 5.
In Figs. \ref{e0_cont} and \ref{i0_cont}, we show detailed dependence on 
$i_*$ as well as $\omega_*$. 
In the inner region, $e = e_0 (a/D)^{5/2}$ and $i = i_0 (a/D)^{3/2}$. 
We investigate the dependence of $e_0$ and $i_0$ on $\omega_*$ and $i_*$, 
in the case with $e_*=1$ and $M_* =1$. 
According to the analytical results 
(Eqs. (\ref{eq:e}) and (\ref{eq:i}) in section 5), 
we scale  
 $e_0$ and $i_0$ by $(15\pi/32 \sqrt{2})(1/ \sqrt{2})$ 
and $(3\pi/8\sqrt{2})(1/ \sqrt{2})$, respectively. 
We denote the scaled $e_0$ and $i_0$ by 
$\tilde{e}_0$ and $\tilde{i}_0$. 
Figure \ref{e0_cont}a shows 
contours of numerically obtained $\tilde{e}_0$ 
as a function of $i_*$ and $\omega_*$. 
The analytical result (Eq. \ref{eq:e}) is plotted in Fig. \ref{e0_cont}b. 
We find dependence on $\omega_*$ is weaker than that on $i_*$.
Since the analytical result shows that $\tilde{i}_0$ is independent of 
$\omega_*$ (Eq. \ref{eq:e}), we plot $\tilde{i}_0$  as a function of $i_*$ 
in Fig. \ref{i0_cont}. 
Filled circles in Fig. \ref{i0_cont} 
show numerically obtained $\tilde{i}_0$. 
We plot the numerical results of ten different 
$\omega_*$ for each $i_*$ in Fig. \ref{i0_cont}. 
$\tilde{i}_0$ is almost completely independent of $\omega_*$, 
which is consistent with the analytical result. 
Figures \ref{e0_cont} and \ref{i0_cont} show that
 the analytical results perfectly agree with numerical results. 
These figures show 
$\tilde{e}_0$ and $\tilde{i}_0$ are symmetric with respect to 
$i_* = 90^{\circ}$. 
Figure \ref{i0_cont} clearly shows $i$ is proportional to 
$\sin 2 i_*$. 

We next show the range of the power-law inner region is regulated by 
$\Omega\sub{kep}$ and $\Omega_*$ (ILB00 also discussed this issue), where  
$\Omega\sub{kep}$ = $\sqrt{G M_1/a^3}$ and 
$\Omega_*$ = $\sqrt{G(M_1+M_2)(1+e_*)/D^3}$ 
are the Keplerian frequency of a planetesimal with semimajor axis $a$ and 
the angular velocity of the passing star at pericenter. 
Figures \ref{M_e_dep} and \ref{is_dep} show that 
if $\Omega\sub{kep}$ $\ga$ $5 \Omega_*$ 
($a/D \siml 0.2$ in case with $e_*=1$ and $M_*=1$), 
$e$ is in proportion to $(a/D)^{2.5}$. 
If $\Omega_*$ $\la$ $\Omega\sub{kep}$ $\la$ $5\Omega_*$ 
($0.2 \siml a/D \siml 0.3 $ in case with $e_*=1$ and $M_*=1$), 
$e$ is pumped up most steeply. 
In this region, 
resonant interactions are important \citep{ost94}.
The condition of $n:1$ commensurability is 
$\Omega_*/\Omega\sub{kep} = (a/D)^{3/2} \sqrt{ (1 + M_*)(1 + e_*)} = 1/n$. 
For example, the 5:1, 4:1 and 3:1 resonances are 
at $a/D \simeq$ 0.21, 0.25 and 0.30 
in the case with 
$e_*=1$, $M_*=1$ and $0^{\circ} < i_* < 90^{\circ}$ 
(the prograde encounters; Figs. \ref{is_dep}a to d). 
The resonances lower than 5:1 dominate non-resonant effects in this case. 
The 1:1, 2:1 and 3:1 are at $a/D$ $\simeq$ 0.63, 0.40 and 0.30 
in the retrograde encounter (Fig. \ref{is_dep}e). 
In the retrograde cases, resonances occur with particles in the 
far side of the disk, so that their effects are relatively weak. 
Numerical simulations suggest that 
the boundary between the inner and outer regions is at the 5:1 
commensurability, $a/D$ $\simeq$ $[(1+M_*)(1+e_*)]^{-1/3}(1/5)^{2/3}$, 
in the case of prograde encounters and at the 3:1, 
$a/D$ $\simeq$ $[(1+M_*)(1+e_*)]^{-1/3}(1/3)^{2/3}$, in the case of 
retrograde encounters. \citet{ost94} analytically 
derived consistent conditions. 
If $\Omega\sub{kep} \la \Omega_*$
($a/D \simg 0.63 $ in case with $e_*=1$ and $M_*=1$), 
there is no Lindblad resonance, so that $a$-gradient is less steep 
than that in the resonant region. 

In the outer and outermost regions, pumped-up $e$ and $i$ 
depend not only 
on initial radial position but also on azimuthal position 
of planetesimals. 
In the case of a prograde encounter, 
particles in the near side of the disk at 
pericenter passage are affected by resonances while ones in the far side 
are hardly affected. 
In these region, $\Omega\sub{kep}$ is not large enough compared with 
$\Omega_*$, so that such asymmetry remains. 
In the case of a retrograde encounter, only far-side planetesimals 
are affected. 
Note that particles with very small $e$ and $i$ exist at resonant points 
depending on the azimuthal position, 
which might be able start runaway accretion (section 6).
 
The change in semimajor axis, 
$\Delta a/a$, is much smaller than that in $e$ and $i$ (in radian) 
in the inner region, as shown by comparison of Fig. \ref{M_e_dep}a 
and Fig. \ref{a0_vs_a}. 
In the inner region, $\Omega\sub{kep}$ $\gg$ $\Omega_*$, so that for particles 
there gravitational potential of the passing star is quasi-stationary. 
In the stationary potential, energy and hence $a$ is conserved. 
Since the potential is not axisymmetric in the heliocentric frame, 
$e$ is changed. Since it is inclined from the orbital plane of the particles, 
$i$ is also changed. 
In the outer and outermost regions with 
$\Omega_*$ $\sim$ $\Omega\sub{kep}$, change in 
$\Delta a/a$ can not be neglected compared with that in $e$ or $i$. 

A stellar encounter can cause disk truncation. 
Many particle are unbound ($e \simg 1$) at $a/D \simg 0.3$ 
after the prograde encounter, 
in the case with $M_*=1$, $e_*=1$ (Fig. \ref{time_d}c).
Some particles are captured by the passing star during a prograde encounter. 
Their eccentricities are usually close to unity 
in the frame in which the passing star is at center. 
(If the passing star also has a planetesimal disk, some planetesimals 
are captured by the primary star.) 
Note that exact amount of the disk truncation 
and the capture depends on the radius of outer edge of the disk 
relative to $D$. 
Such strong encounters would have many interesting features as mentioned in 
Section 1. However, as shown in section 2, in order to discuss the effects 
on planet accretion, the regimes of $e$, $i$ $\sim$ 0.01 are important. 
The pumped-up $e$ and $i$ in such regimes are near the inner region. 
Hence, analytical linear calculations would well predict the effects on 
planet accretion.

\section{ANALYTICAL CALCULATION}

We derive analytical formulae of pumped-up $e$ and $i$
in the inner region. 
Orbital integrations show that in that region, 
$e = e_0 (a/D)^{5/2}$ and $i = i_0 (a/D)^{3/2}$. 
The analytical formulae explain these dependence 
on $(a/D)$ as well as dependence of $e_0$ and $i_0$
on $M_*$, $i_*$, $\omega_*$ and $e_*$. 
\citet{heggie96} derived change in eccentricity of binary stars 
which encounter a passing single star, using the Laplace-Runge-Lenz vector. 
Using the Gauss's equations, we derive change in inclination 
as well as eccentricity. 
In the limit of null mass of a binary companion, the formula by 
\citet{heggie96} is equivalent to out formula for eccentricity. 
The analytical results show the phase alignment, so that $e$ and $i$ 
do not necessarily determine relative velocity between perturbed 
planetesimals. 
We numerically study the degree of the phase alignment in section 6. 

We are mostly concerned with parabolic encounters ($e_*=1$), 
since for encounters we are interested in, $e_*$ is not far from 
1, as follows. 
(We also calculate hyperbolic encounters ($e_* > 1$)with the same 
analytic method in Appendix.) 
The specific angular momentum ($l_*$) and 
specific energy ($E_*$) of the passing star orbit 
relative to the primary star are $\sim$ $v\sub{*} D$ $\sim$ 
$\sqrt{v\sub{*d}^2+ [v\sub{kep}(D)]^2 } \times D$ and 
$v\sub{*d}^2/2$, respectively, where $v\sub{*d}$ is  
 velocity dispersion of stars in a cluster. 
Thereby, $e_*$ $= \sqrt{1+2l_*^2E_*/G^2(M_1+M_2)^2}$ $\sim$ 
$\sqrt{1+ (v\sub{*d}/v\sub{kep})^2[1+(v\sub{*d}/v\sub{kep})]^2}$. 
Since $v\sub{*d}$ is $\sim 1$ km/s \citep{binney1987} and 
$v\sub{kep}(D)$ is larger than  1 km/s for $D \la$  1000AU, 
which we are interested in, $e_*$ is $\sim 1$. 

We adopt the following approximations 
to derive pumped-up $e$ and $i$ in the inner region.
\begin{itemize}
\item[(i)] $\Delta a/a$ is neglected, since $\Delta a/a \ll e, i$, 
\item[(ii)] orbital averaging is applied for planetesimal orbits, since 
$\Omega\sub{kep} \gg \Omega_*$ \citep{larwood97}, 
\item[(iii)] $e$, $i$, $a/D$ $\ll$ 1. 
\end{itemize}

In the equation of motion of a planetesimal given by Eq. (\ref{eq:eq_motion}), 
the second and third terms in the r. h. s. stand for the perturbation forces 
of the passing star. We define 
\begin{equation}
{\bf F}\sub{perturb}= \frac{GM_2}{\mid {\bf R}-{\bf r} \mid^3}
({\bf R}-{\bf r}) - \frac{G M_2}{\mid {\bf R} \mid^3}
{\bf R}.
\label{eq:perturbation}
\end{equation}
We divide the perturbation force into $r$, $\theta$, and $z$ components, 
\begin{equation}
{\bf F}\sub{perturb}= \bar{R} {\bf e}_r + \bar{T} {\bf e}_{\theta} 
+ \bar{N} {\bf e}_z,
\label{eq:perturbation2}
\end{equation}
where ${\bf e}_r$ ($= {\bf r}/\mid {\bf r} \mid$), 
${\bf e}_{\theta}$ and ${\bf e}_z$ 
are unit vectors in the radial, 
tangential, and normal components in initial orbital plane of 
the planetesimal disk, respectively. 
We define $h = e \sin ( \omega + \Omega),
k = e \cos (\omega + \Omega),
p = \sin i \sin \Omega$ and
$q = \sin i \cos \Omega$, 
where $\omega$ and $\Omega$ are the argument of pericenter and longitude of 
ascending node of a planetesimal. 
The equations of motion with these orbital elements in $h,k,p,q$ $\ll$ 1 
($e, i$ $\ll$ 1: assumption (iii)), which are called Gauss's equations,
 are \citep{brouwer}
\begin{eqnarray}
\frac{d h}{dt} &\simeq& \sqrt{\frac{a}{G M_1}} ( - \bar{R} \cos \theta
+ 2 \bar{T} \sin \theta ),
\label{eq:eq_gauss1}\\
\frac{d k}{dt} &\simeq& \sqrt{\frac{a}{G M_1}} (  \bar{R} \sin \theta
+ 2 \bar{T} \cos \theta ),
\label{eq:eq_gauss2}\\
\frac{d p}{dt}& \simeq & \sqrt{\frac{a}{G M_1}} \bar{N} \sin \theta ,
\label{eq:eq_gauss3}\\
\frac{d q}{dt} & \simeq& \sqrt{\frac{a}{G M_1}} \bar{N} \cos \theta ,
\label{eq:eq_gauss4}
\end{eqnarray}
where 
$\theta = f + \omega + \Omega$
($f$ is true anomaly) and we retained the lowest order terms of $e$ and $i$ in
the right hand side. 

We expand $\bar{R}$, $\bar{T}$, $\bar{N}$ in terms of
 $a/D$ up to the terms of $(a/D)^2$ 
(see Appendix, Eqs.(\ref{eq:R}), (\ref{eq:T}), and (\ref{eq:N})).
Corresponding to assumption (ii), we take orbital averaging, e.g.,
 \begin{equation}
 \left< \frac{d h}{d t} \right> = \frac{\int^{2\pi}_{0} (dh/dt) d \theta}{ 2 \pi}.
\end{equation}
The averaged $dh/dt$, $dk/dt$, $dp/dt$ and $dq/dt$ are shown in Eqs. 
(\ref{eq:ave1}), (\ref{eq:ave2}), (\ref{eq:ave3}) and (\ref{eq:ave4}) 
in Appendix.
Finally, we integrate them with the stellar passage, assuming constant $a$ 
of planetesimals (assumption (i)) to obtain changes in
 $h$, $k$, $p$, and $q$ caused by the stellar passage in parabolic orbit 
 (Appendix, Eqs. (\ref{eq:app_h_e}) to (\ref{eq:app_q_e})). 

Since we start with $h, k, p, q = 0$, their changes are equal to final $h$,
$k$, $p$,
 and $q$ of planetesimals.
Since $e = \sqrt{ h^2 + k^2}$ and $i = \sqrt{p^2 + q^2}$, we obtain
 final $e$ and $i$ as
 \begin{eqnarray}
 e &\simeq& \frac{15 \pi}{32 \sqrt{2}}
\frac{M_*}{\sqrt{1+M_*}}
 \left( \frac{a}{D} \right)^{\frac{5}{2}}
 \nonumber
 \\ && \times \left[ \cos^2 \omega_*
 \left( 1- \frac{5}{4} \sin^2 i_*  \right)^2  + \sin^2 \omega_* \cos^2 i_*
 \left( 1 - \frac{15}{4} \sin^2 i_* \right)^2 \right]^{\frac{1}{2}},
\label{eq:e} 
 \\
 i &\simeq& \frac{3 \pi}{8 \sqrt{2}} \frac{M_*}{\sqrt{1+M_*}}
 \left( \frac{a}{D} \right)^{\frac{3}{2}} |\sin 2 i_* |, 
 \label{eq:i}\\
\tilde \omega & \simeq & \arctan \left[ -
	\frac{\cos \omega_* (5 \cos^2 i_* -1)}	 
	{\sin \omega_* \cos i_* (15 \cos^2 i_* -11)} \right],
\label{eq:t_omega}\\
\Omega &\simeq& \left\{ \begin{array}{c}
	\pi/2 \\ - \pi/2
       \end{array} \right.
\hspace{2cm}
\begin{array}{c}
\mbox{for} \\ \mbox{for} 
       \end{array} 
\hspace{0.5cm}
\begin{array}{c}
0^{\circ} \leqq i_* \leqq 90^{\circ}, \\ 
	 90^{\circ} < i_* \leqq 180^{\circ},
       \end{array} 
\label{eq:Omega}
 \end{eqnarray}
where ${\tilde \omega}$ is the longitude of pericenter and 
$= \omega + \Omega$. 
The detailed derivation is shown in Appendix. 
We have analytically explained the power-law dependence of pumped-up $e$ and $i$ on $(a/D)$ in inner region.
Since the orbital averaging is valid, $\tilde{\omega}$ and $\Omega$ of 
the perturbed orbit do not depend on azimuthal position of a planetesimal as 
Eqs. (\ref{eq:t_omega}) and (\ref{eq:Omega}). Such phase alignment of 
planetesimals' orbits, leads to reduction of the colliding velocity 
between planetesimals (section 6). 

We plot the analytical formulae (\ref{eq:e}) and (\ref{eq:i})
 with dashed lines with triangles in Figs. \ref{M_e_dep} and \ref{is_dep}. 
$e_0$ ($=e/(a/D)^{5/2}$) and $i_0$ ($=i/(a/D)^{3/2}$) 
of the analytical formulae are showed in Figs. \ref{e0_cont}b and 
\ref{i0_cont}. 
The analytical formulae show excellent agreement with the numerical results
 in the inner power-law inner region.
We examined the agreement in the cases with other parameters and found that
 the difference is always less than a factor 2. 

The pumped-up $e$ and $i$ depend on phase angles $i_*$ and $\omega_*$ of a 
stellar encounter. 
The average 
of $e$ and $i$ with $\omega_*$ and $i_*$ is 
\begin{eqnarray}
e\sub{ave} = \left[ \frac{1}{4 \pi} 
\int^{2 \pi}_{0} d \omega_* \int^{\pi}_{0} d i_* (\sin i_*) e^2 
\right]^{\frac{1}{2}}
= \frac{15 \pi}{32 \sqrt{7}}
\frac{M_*}{\sqrt{1+M_*}}
 \left( \frac{a}{D} \right)^{\frac{5}{2}},
\label{eq:e_ave}\\
i\sub{ave} = \left[ \frac{1}{4 \pi} 
\int^{2 \pi}_{0} d \omega_* \int^{\pi}_{0} d i_* (\sin i_*) i^2 
\right]^{\frac{1}{2}}
= \frac{\sqrt{15} \pi}{20}
\frac{M_*}{\sqrt{M_*+1}}\left( \frac{a}{D} \right)^{\frac{3}{2}}.
\end{eqnarray}
We use $e\sub{ave}$ and $i\sub{ave}$
to estimate the boundary of planet accretion in section 7.

\citet{ost94} analytically investigated 
the change in energy ($\Delta E\sub{disk}$) 
and perpendicular angular momentum ($\Delta L_{{\rm disk}, z}$) 
of an entire circumstellar disk during a parabolic stellar encounter. 
Although the author was concerned mainly with resonant region for 
$r\sub{disk,max}$ $\ga$ $0.2 D$, 
 the author presented phase-averaged 
approximate $\Delta L_{{\rm disk}, z}$ 
(but not $\Delta E\sub{disk}$) for $r\sub{disk,max}$ $\la$ $0.2 D$ 
(non-resonant region). 
Since specific perpendicular angular momentum $\Delta L_z$ 
are written by $L_0 (\Delta a/a $ 
$- \Delta e^2$ $- \Delta i^2)/2$ where $L_0$ is the initial angular momentum, 
$\Delta L_{{\rm disk}, z}$ is obtained by 
integrating $\Delta L_z$ over radius from $r\sub{disk,min}$ to 
$r\sub{disk,max}$, 
using our results in the power-law inner region. 
Our $\Delta L_{{\rm disk}, z}$ is larger by a factor $\sim {\cal O}(10)$ 
than the author's with less steep $a$-gradient. 

\citet{heggie96} derived formula of change in eccentricity of binary stars 
after a single passing encounter. 
To derive this formula, they use the Laplace-Runge-Lenz vector. 
Their formula of change in eccentricity of binary stars with circular orbit 
is perfectly the same as our formula derived by a different method, 
in limit of null mass of a binary companion.

\section{RELATIVE VELOCITY}

As shown in the last section, difference of $\tilde{\omega}$ or $\Omega$ 
between planetesimals 
are almost 0 as long as the linear analysis is valid. 
We investigate the relative velocity between planetesimals 
after a stellar passing, using the numerical data.
Figure \ref{e0_omega} show that $\tilde{\omega}$ and $\Omega$ after 
the stellar encounter with ($i_*$, $\omega_*$, $M_*$, $e_*$) = ($5^{\circ}$, 
$90^{\circ}$, 1, 1). 
$\tilde{\omega}$ and $\Omega$ are aligned in the region 
where $e$ and $i$ have the power law radial dependence. 
In the outer region where the resonances are important, 
$\tilde{\omega}$ is randomly distributed 
(Retrograde encounters show similar results). 

\citet{whitmire} evaluated 
relative velocities between planetesimals 
whose orbits are crossing assuming coplanar orbits. 
Extending their method to three-dimension with assumption $i \la 1$(radian), 
we evaluated the relative velocity between planetesimals 1 and 2. 
For $i \la 1$, the projective orbit of a planetesimal on the initial disk is 
almost the same as the case with $i=0$. 
We calculate the relative velocity, if 
$| z_1$ $ - z_2 |$ is smaller than the sum of planetesimal radii 
at the point where the projective 
orbits are crossing, where $z_{1(2)}$ is 
position vertical to the initial disk. 
The component on the initial disk $v_{0,xy}$ and 
that vertical to the disk $v_{0,z}$ at a crossing point are 
\begin{eqnarray}
v_{0,xy}^2 &=& GM_1 \left[ \frac{e_1 \sin ( f_2 - \Delta \tilde{\omega})}{\sqrt{p_1}}
	- \frac{e_2 \sin f_2 }{\sqrt{p_2}} \right]^2 
	+ \left[ \frac{\sqrt{p_1}}{r}-\frac{ \sqrt{p_2} }{r} \right]^2,
\label{eq:ur}\\
v_{0,z} &=& \sqrt{GM_1} \left| \frac{i_1 [ \cos ( f_2 - \Delta \tilde{\omega} 
				+ \omega_1)+
	e_1 \cos \omega_1]}
	{\sqrt{p_1}}
 	- \frac{i_2 [ \cos ( f_2 + \omega_2) + e_2 \cos \omega_2]}{\sqrt{p_2}}
	\right|,
\label{eq:uz}
\end{eqnarray}
where $p_{1(2)} = a_{1(2)} ( 1 - e_{1(2)})$, $\Delta \tilde{\omega}$ 
and $\Delta \Omega$ are $\tilde{\omega}_2 - \tilde{\omega}_1$ and 
$\Omega_2 - \Omega_1$. 
$f_2$ is true anomaly of planetesimal 2 at the crossing point, 
satisfying 
\begin{equation}
\cos f_2 = \frac{-A B \pm C \sqrt{ C^2 + B^2 - A^2}}{B^2 + C^2}
\end{equation} 
where
\begin{eqnarray}
A &=& p_2-p_1,\\
B &=& e_1 p_2 \cos \Delta \tilde{\omega} - e_2 p_1,\\
C &=& e_1 p_2 \sin \Delta \tilde{\omega}.
\end{eqnarray}
$r$ is radial position of planetesimals at the crossing point, 
given by 
\begin{equation}
r = \frac{P_2}{1+ e_2 \cos f_2}.
\end{equation}

In Fig. \ref{e0_rel}, $e$ and $i$ are compared with 
$e\sub{eff} = v_{0,xy}/v\sub{kep}$ and $i\sub{eff} = v_{0,z}/v\sub{kep}$, 
which are calculated through 
Eqs. (\ref{eq:ur}) and (\ref{eq:uz}) 
with the data in Fig. \ref{e0_omega}. 
We also plot 
\begin{eqnarray}
e\sub{rel} &\equiv& | {\bf e}_2 - {\bf e}_1 | = e_1^2 +e_2^2 - 2 e_1 e_2 \cos \Delta \omega, \\
i\sub{rel} &\equiv& | {\bf i}_2 - {\bf i}_1 | = i_1^2 +i_2^2 - 2 i_1 i_2 \cos \Delta \Omega, 
\end{eqnarray}
where  ${\bf e}_j$ ${\bf i}_j$ ($j =$ 1,2) are 
$(e_j \cos \tilde{\omega}_j, e_j \sin \tilde{\omega}_j)$ and 
$(i_j \cos \Omega_j, i_j \sin \Omega_j)$. 
In Fig. \ref{e0_rel}, we found $e\sub{rel} \simeq e\sub{eff}$ and 
$i\sub{rel} \simeq i\sub{eff}$. 
The effect of the phase alignment is more clear 
with $e\sub{rel}$ and $i\sub{rel}$ rather than with 
$e\sub{eff}$ and $i\sub{eff}$, 
because $e\sub{rel}$ and $i\sub{rel}$ are directly related with 
$\Delta \tilde{\omega}$ and $\Delta \Omega$. 
A stellar encounter results in alignment of $\tilde{\omega}$ and $\Omega$ 
($\Delta \tilde{\omega}$ $\simeq 0$ and $\Delta \Omega$ $\simeq 0$) 
in inner region, $e\sub{eff}$ ($e\sub{rel}$) $\ll e$ and 
$i\sub{eff}$ ($i\sub{rel}$) $\ll i$. 
In outer region, the linear approximation breaks down and 
$\Delta \tilde{\omega}$ and $\Delta \Omega$ are randomly distributed, so 
that $e\sub{rel}$ $\simeq e$ and $i\sub{rel}$ $\simeq i$. 
Figure \ref{e0_omega} shows that the alignment breaks down at $a/D$ 
$\sim$ 0.2 for $\tilde{\omega}$, and $\sim$ 0.3 for $\Omega$, 
resulting in more rapid  increase in $e\sub{rel}$ with $a$ than $i\sub{rel}$. 
As a result, at $e\sub{eff}$ $\sim$ $e\sub{eff,crit}$ 
$\sim$ 0.01, 
$e\sub{eff}$ $\sim$ $e$ $> i\sub{eff}$. 
Also for other encounter parameters, we find a similar trend. 
We thus estimate planet formation region, 
using Eqs. (\ref{eq:e_bound}) and (\ref{eq:e}) with $e\sub{eff}$ $\sim e$. 

In Fig. \ref{e0_rel}, we also find that at a resonant point 
($a/D \simeq$ 0.2), 
all of $e$ and $e\sub{eff}$ have wide variety of values caused by the resonant 
feature. 
Although many planetesimals have $e\sub{eff}$ well beyond $e\sub{eff,crit}$, 
but some fraction has $e\sub{eff} \ll e\sub{eff,crit}$ should be very fast. 
If it can precede disruption due to collisions with high $e\sub{eff}$ 
planetesimals to form large bodies with high $e\sub{eff,crit}$ quickly, 
planet can be formed at this particular resonant point. 

We also study the cases with non-zero initial $e$ and $i$ 
because self-gravity of planetesimals or more distant stellar encounters 
would occur before the strongest encounter. 
Figure \ref{e5_omega} shows 
the $e$, $i$, $\tilde{\omega}$ and $\Omega$ of 
planetesimals after the encounter 
with initial $e$ and $i$ $ = 1 \times 10^{-4}$. 
$\tilde{\omega}$ and $\Omega$ are initially at random.
The encounter parameters are ($i_*$, $\omega_*$, $M_*$, $e_*$) 
$=$ ($5^{\circ}$, $90^{\circ}$, 1,1). 
If change of $e$ and $i$ by the encounter are much larger than 
the initial $e$ and $i$, 
the features of $e$ and $i$ are almost the  
same as those in the case of initial $e$ and $i$ = 0; 
$\tilde{\omega}$ and $\Omega$ are aligned 
If the pumped-up $e$ and $i$ are smaller than the initial $e$ and $i$, 
$\tilde{\omega}$ and $\Omega$ remain at random.  

After the stellar encounter, self-gravity of planetesimals, 
collision between them and gas drag would be important for accretion of 
planetesimals on a longer time scale. 
They may keep the phase alignment rather than destroy it 
\citep[e.g.,][]{gt82,marzari00,ito01}, until a strong perturber such as a giant 
planet is formed. 
We need more study about these effects on planetesimal accretion. 
\section{SIZE OF A PLANET FORMING REGION}

In the last section, we have derived analytical expressions
 for $e$ and $i$ pumped up by a passing stellar encounter. 
The expressions perfectly agree with numerical calculations in the
 region $e \la 0.01$.
In this section, we estimate the size of planet forming region
 in a protoplanetary disk, using these analytical expressions.

According to the argument in the last section, 
 the collision between planetesimals 
results in disruption rather than accretion (Eq. (\ref{eq:e_bound})),
\begin{equation}
 e\sub{eff} > \frac{v_{\rm esc}}{v_{\rm kep}} \sim 0.01 \left( \frac{m}{10^{22}
\mbox{g}}
\right)^{\frac{1}{3}} \left( \frac{a}{10 \mbox{AU}} \right)^{\frac{1}{2}},
 \label{eq:e_bnd}
\end{equation}
 where $m$ is mass of a planetesimal.
For $e \ga 0.01$, $\tilde{\omega}$ and $\Omega$ are not aligned and $e$ $\simeq$ $e\sub{eff}$. 

As shown in sections 3 and 4,
 the radial gradient of $e$ is so steep that there is
 a sharp boundary of the disk that
 divides the planet forming region and the disruptive region where planet
 formation is inhibited.
Since $e\sub{eff}$, which includes the effect of the phase alignment, 
has steeper gradient than $e$, 
the actual boundary should be sharper. 
Here we derive the radius of the boundary as a function of
 physical parameters of the stellar cluster which the host star of
 the planetary system belonged to.

In the case with $e \ga 0.01$, the analytical expressions
 slightly underestimate the pumped-up $e$, 
since resonant effects are also important in this case. 
However, we can use
 them to estimate the boundary radius, because
 the radial gradient of $e$ is so steep that the underestimation
 hardly changes the boundary radius. 

In a stellar cluster, $e_*$ $\sim$ 1, as mentioned in Section 5. 
Substituting Eq. (\ref{eq:e_ave}) into Eq. (\ref{eq:e_bound}),
 we obtain the boundary radius $a_{\rm planet}$ as
 \begin{equation}
 a\sub{planet} \sim 40 \left(\frac{m}{10^{22} \mbox{g}}\right)^{\frac{1}{6}}
 \left( \frac{F}{2} \right)^{\frac{1}{4}}
 \left( \frac{D}{150 \mbox{AU}} \right)^{\frac{5}{4}} \mbox{AU},
 \label{eq:a_planet}
\end{equation}
 where $F = M_*+1/M_*^2$. 
If $M_*$ is 1, $F$ is 2. 
The factor $(F/2)^{1/4}$ can not significantly deviate from 1. 
The dependence of $m$ on $a\sub{planet}$ is also very weak. 
Therefore, $a\sub{planet}$ depends almost only on $D$.
A stellar encounter with $D \sim 150$-200AU restricts the disk radius of a
 planetary system (the disk radius of planet forming region) to
 40-60AU. 

Considering evaporation process of a stellar cluster,
 \citet{adams00} estimated effective $D$ before the evaporation as
\begin{equation}
D \sim 200 \left( \frac{R_{\rm cluster}}{2 \mbox{pc}}\right)
 \left(\frac{N}{2000}\right)^{-1} \mbox{AU},
\label{eq:est_D}
\end{equation}
 where $N$ is number of stars in a stellar cluster 
and $R\sub{cluster}$ is size of the cluster. 
For the Trapezium cluster in Orion, $N \sim 2300$ and
 $R_{\rm cluster} \sim 2$ pc.
In a dense cluster like Orion Trapezium,
 $D$ is as small as $200$AU, so that $a\sub{planet} \sim 40$-60AU.
ILB00 demonstrated that
 the high eccentricity and inclination of objects in the outer Kuiper Belt
 may be explained by the stellar encounter with $D \sim 150$-200AU, which
 may suggest the Sun was born in a dense stellar cluster.

So far, we have only considered passing of a single star,
 however, passing of binary stars would also be important.
\citet{lau98} and \citet{adams00} (also see the next section) suggested that
 passing binary encounters are more disruptive than passing
 single-star encounters.
If we take into account the effects of passing binary encounters,
 $a\sub{planet}$ may be smaller than Eq. (\ref{eq:a_planet}). 

We should investigate distribution of $D$ (not only an effective
 value) as well as the effects of passing binary encounters,
 to discuss diversity of sizes of planetary systems in more detail.

Note that as discussed in the last section, 
less than planetesimals might 
coalesce with each other, resulting in a kind of runaway growth; 
planets might be formed at particular resonant location 
beyond $a\sub{planet}$. 

\section{CONCLUSION \& DISCUSSION}

We have investigated the effects of a passing stellar encounter 
 on a planetesimal disk 
 through orbital integration and analytical calculations.
Since stars are generally born as members of a cluster
 and would stay in the cluster on timescales 
 more than $10^8$ years \citep{kroupa1995,kroupa1998},
 a relatively close encounter, e.g., one with distance $\sim$
 200AU is likely to occur during formation age of
 a planetary system.

We considered that a disk of massless particles
 (planetesimals) orbiting a primary star encounters a passing single star. 
Encounter parameters are pericenter distance of the encounter ($D$), 
 the argument of perihelion ($\omega_*$), eccentricity ($e_*$) 
 and inclination ($i_*$) of the orbit of the passing star,
 and the mass ratio ($M_*$) of the passing star's mass to the primary one. 
We showed that the pumped-up orbital eccentricities $e$ and
 inclinations $i$ of planetesimals have steep radial gradient.
In the inner region at semimajor axis $a \la $ $\alpha$ 
$[(1+M_*)/2]^{1/3}$ $[(1+e_*)/2]^{1/3}$ 
$D$, ($\alpha$ $\simeq$ 0.2 for a prograde encounter and $\alpha$ $\sim$ 0.3 
for a retrograde encounter) 
$e \propto (a/D)^{5/2}$ and $i \propto (a/D)^{3/2}$, 
 independent of the encounter parameters.
The result is also independent of initial azimuthal position of planetesimals,
 because orbital period of planetesimals is much shorter than
 passing time scale.

In the outer region $a \ga$ $\alpha$ 
$[(1+M_*)/2]^{1/3}$ $[(1+e_*)/2]^{1/3}$ $D$, 
the radial gradient is steeper,
 but is not expressed by a single power-law.
In this region, resonant effects are more important \citep{ost94}.

The longitude of pericenter $\tilde{\omega}$ and 
ascending node $\Omega$ are aligned between planetesimals, in power-law region of $e$ and $i$ (Fig. \ref{e0_omega}). 
We investigated relative velocity 
between planetesimals whose orbits cross. 
We define $e\sub{eff} = v_{0,xy}/v\sub{kep}$ and 
$i\sub{eff} = v_{0,z}/v_{kep}$ where $v_{0,xy}$ and $v_{0,z}$ are planar 
and vertical components of the relative velocity. 
We found that $e\sub{eff}$ and $i\sub{eff}$ are smaller than $e$ and $i$ 
by many orders of magnitude. 
In outer region, the phase alignment brake down, 
so that $e \sim e\sub{eff}$ and $i \sim i\sub{eff}$. 

The stellar perturbations significantly affect
 the outer part of a planetesimal disk.
Even in the inner disk, $e\sub{eff}$ and $i\sub{eff}$ 
that are pumped up to $\ga 0.01$ inhibit further planetesimal accretion, 
because the relative velocity corresponding to 
$e\sub{eff}, i\sub{eff} \ga 0.01$ 
 exceeds the surface escape velocity of planetesimals and collisions between 
 planetesimals should be disruptive. 
We investigate the relatively weakly perturbed region 
 with $e, i \sim 0.01$ to study the effects on planetesimal accretion.
We have derived analytical expressions of $e$ and $i$ in the power-law 
inner region. 

With the approximations of constant the semimajor axis $a$ and
 orbit averaging of
 planetesimals' motion, 
we derived analytical expressions
 of the pumped-up $e$ and $i$ (Eqs. (\ref{eq:e}) and (\ref{eq:i})) 
in the case with a parabolic encounter ($e_* = 1$)
 as
\begin{eqnarray}
e &\simeq& \frac{15 \pi}{32 \sqrt{2}} \frac{M_*}{\sqrt{M_*+1}}
 \left( \frac{a}{D} \right)^{5/2} \nonumber
\\ && \times \left[ \cos^2 \omega_* \left( 1-\frac{5}{4} \sin^2 i_* \right)^2
+ \sin^2 \omega_* \cos^2 i_* \left( 1 - \frac{15}{4} \sin^2 i_* \right)^2 
\right]^{1/2},\\
i &\simeq& \frac{3 \pi}{8 \sqrt{2}} \frac{M_*}{\sqrt{M_*+1}}
\left( \frac{a}{D} \right)^{3/2} | \sin 2 i_* |.
\end{eqnarray}
Note that the dependence on $\omega_*$ is weak in both expressions 
and that on $i_*$ is also weak in the former expression 
(Figs. \ref{e0_cont} and \ref{i0_cont}). 
The analytical formulae agree with the numerical results
 within a factor 2. 
We also calculated $e$ and $i$ in the case with a hyperbolic 
encounter ($e_* > 1$) in Appendix. 

At $e\sub{eff} \sim 0.01$, $e \sim e\sub{eff}$ and $e\sub{eff}$ is larger than $i\sub{eff}$, because the phase alignment for $\tilde{\omega}$ breaks down 
at smaller $a$ than that for $\Omega$. 
Hence the relative velocity is regulated by $e$ at $e\sub{eff} \sim$ 0.01 and 
disruptive collisions $e \ga e\sub{eff,crit}$ where $e\sub{eff,crit}$ is given as 
Eq. (\ref{eq:e_bound}). 
Since the radial gradient of $e$ and $e\sub{eff}$ are so steep that there is
 a sharp boundary in the disk that
 divides the planet forming region and the disruptive region where planet
 formation is inhibited. 
Planetesimal orbits are significantly modified beyond the boundary,
 while they are almost intact inside the boundary.
We derived the boundary radius ($a\sub{planet}$) by
 the condition that the pumped-up velocity dispersion of planetesimals
 is equal to their surface escape velocity (Eq.(\ref{eq:a_planet})): 
 \begin{equation}
 a_{\rm planet} \sim 40 \left(\frac{m}{10^{22} \mbox{g}}\right)^{\frac{1}{6}}
 \left( \frac{F}{2} \right)^{\frac{1}{4}}
 \left( \frac{D}{150 \mbox{AU}} \right)^{\frac{5}{4}} \mbox{AU},
\end{equation}
 where $F \simeq M_*+1/M_*^2$ and $m$ is planetesimal mass.
The variation of $(F/2)^{1/4}$ is small. 
The dependence of $a_{\rm planet}$ on $m$ is very weak.
Thereby $a_{\rm planet}$ depends almost only on $D$.
In a dense cluster like Orion Trapezium,
 $D$ is as small as $200$AU \citep{adams00}.
Such a stellar encounter with 150-200AU restricts the disk radius of a
 planetary system (the disk radius of planet forming region) to
 40-60AU.

In the Solar system, there would be no planetary-sized object
 beyond Neptune at 30AU.
Kuiper-belt objects beyond Neptune
 have velocity dispersion considerably larger than their
 surface escape velocity.
These might be accounted for by a stellar encounter in
 a dense stellar 
(ILB00; Adams \& Laughlin 2001). 
Our Solar system may have belonged to a dense stellar
 in the formation age.

In a dense stellar cluster, planetary systems cannot be 
 significantly larger than the size of the planetary region of
 our Solar system ($\sim 30$-40AU).
To discuss diversity of sizes of planetary systems in detail, 
 we will need to investigate distribution of $D$ (not only an effective
 value).
In the present paper, we have only considered passing of a single star,
 however, passing of binary stars would also be important.
\citet{lau98, lau00} suggested that
 passing binary encounters are more disruptive than passing
 single-star encounters.
We performed several runs of passing close binary stars. 
If the total binary mass is equal to a single passing star mass, 
the results show that the encounter of close binary stars is 
similar to the single stellar encounter, except that $e$ 
is pumped up more highly at some resonant positions corresponding to the 
binary frequency 
even in the inner region. 
As long as close binary cases are considered, 
 $a_{\rm planet}$ may be estimated with the single star's equation using 
binary total mass. 

The effects of cumulative distant (large $D$) encounters may be neglected 
compared with a few closest encounters. The successive distant encounters 
change $h, k, p, q$ defined in section 5 constructively or destructively, 
so that their averages would be zero. Their dispersions depend on $D$ with 
large negative power-indexes ($< -3$), so that the cumulative effects 
integrated by $2\pi D d D$ would quickly vanish with $D$. 

Circumstellar dust disks around stars are observed.
 This dust may be secondary, 
because dust would be removed before stellar age, 
by radiative pressure and pointing Robertson effect
\citep[e.g.,][]{artymo97,backman95}. 
Disruptive collisions at $a > a_{\rm planet}$ would continuously produce dust
 materials, which might form dust-debris disks around Vega-like stars.
On the other hand, planetesimals grow to be large bodies 
at $a < a_{\rm planet}$, not producing dust materials. 
Although radiative pressure and pointing Robertson effect
 would cause migration of the produced dust materials 
\citep[e.g.,][]{artymo97,backman95},
 $a_{\rm planet}$ correspond to 
 the radius of inner holes of dust-debris disks. 
The radii of the observed inner holes around 
 $\epsilon$ Eridani \citep{greav98},
 HD 14156 \citep{andrillat90,sahu98,aug99,wei99,wei00}, 
HD 207129 \citep{jour99} and HR 4796A 
\citep{jura95,faj98,jaya98} are 30-100AU, which are comparable
 to $a_{\rm planet}$ in the case of relatively dense clusters.
We need more detailed analysis of distribution of $a_{\rm planet}$ 
 as a function of environment parameters of stellar clusters,
 combined with discussions of collision outcomes and migration
 processes of dust materials, to discuss the diversity of inner holes of
 dust-debris disks.

\section*{Appendix}

We integrate Eqs. (\ref{eq:eq_gauss1}), (\ref{eq:eq_gauss2}),
 (\ref{eq:eq_gauss3})
and (\ref{eq:eq_gauss4}) and derive expressions of
$e$ and $i$ pumped up by a passing star as functions of
$a/D$, $M_*$, $\omega_*$, $e_*$ and $i_*$.

${\bf F}_{\rm perturb}$ defined by Eq. (\ref{eq:perturbation})
 is expanded as a series of $a/D$ up to the order $(a/D)^2$:
 \begin{equation}
 {\bf F}_{\rm perturb} \simeq \frac{GM_2}{R^3}
 \left[\left(3 \frac{ {\bf R} \cdot {\bf e}_r}{R} 
{\bf R} - R {\bf e}_r \right)\left(\frac{a}{R} \right)+
 \left\{ \left( \frac{15}{2} \frac{({\bf R} \cdot {\bf e}_r )^2}{R^2}
-\frac{3}{2} \right) {\bf R} 
- 3 ( {\bf R} \cdot {\bf e}_r) {\bf e}_r \right\}  
\left(\frac{a}{R} \right)^2
 \right],
 \label{eq:perturb2}
 \end{equation}
 where $R = \mid {\bf R} \mid$ and ${\bf e}_r = {\bf r}/a$. 
We denote ${\bf F}_{\rm perturb} =$ ${\bar R} {\bf e}_r$ 
$+ {\bar T} {\bf e}_{\theta}$ $+ {\bar N} {\bf e}_z$, where 
${\bf e}_{\theta}$ and ${\bf e}_z$ are unit vectors in the 
tangential and vertical directions. 
We denote that ${\bf e}_r = (\cos \theta, \sin \theta, 0)$ and
 ${\bf R} = (R_x, R_y, R_z)$ 
in the Cartesian coordinates where the initial planetesimal disk is 
on the $x$-$y$ plane and the $x$-axis is directed to ascending node of 
the passing star's orbit.  
The components, $\bar{R}, \bar{T}$ and $\bar{N}$ are
\begin{eqnarray}
\bar{R} &\simeq& \frac{GM_2}{R^3} 
\left[\left(3 \frac{({\bf R} \cdot {\bf e}_r)^2}{R}
- R \right)\left(\frac{a}{R} \right) 
+\left\{ \left( \frac{15}{2} \frac{({\bf R} \cdot {\bf e}_r)^2}{R^2}
-\frac{9}{2}\right) 
{\bf R} \cdot {\bf e}_r
\right\}\left(\frac{a}{R} \right)^2 \right], 
 \label{eq:R}\\
\bar{T} &\simeq& \frac{GM_2}{R^3} 
\left[ 3 \frac{({\bf R} \cdot {\bf e}_r)}{R}
\phi 
\left(\frac{a}{R} \right)  
\left\{ \left( \frac{15}{2} \frac{({\bf R} \cdot {\bf e}_r)^2}{R^2}
-\frac{3}{2} \right) \phi \right\}
\left(\frac{a}{R} \right)^2 \right], 
 \label{eq:T}\\
\bar{N} &\simeq& \frac{GM_2}{R^3} 
\left[ 3 \frac{({\bf R} \cdot {\bf e}_r)}{R} R_z
\left(\frac{a}{R} \right) 
+\left( \frac{15}{2} \frac{({\bf R} \cdot {\bf e}_r)^2}{R^2}
-\frac{3}{2}\right)R_z
\left(\frac{a}{R} \right)^2 \right],
\label{eq:N}
\end{eqnarray}
where $\phi$ is $z$ component of ${\bf e}_r \times {\bf R}$. 

Substituting Eqs. (\ref{eq:R}), (\ref{eq:N}), and (\ref{eq:T}) 
into Eqs. (\ref{eq:eq_gauss1}), (\ref{eq:eq_gauss2}),
 (\ref{eq:eq_gauss3}) and (\ref{eq:eq_gauss4}) 
and taking the orbit averaging, e.g.,
 \begin{equation}
 \left< \frac{dh}{dt} \right> 
= \frac{\int^{2\pi}_{0} (dh/dt)  d \theta}{ 2 \pi}, 
\end{equation}
where ``$< \hspace{1em} >$'' means the orbit averaging, 
we obtain 
\begin{eqnarray}
\left< \frac{dh}{dt} \right> &\simeq& - \frac{GM_2}{R^3}
\sqrt{\frac{a}{GM_1}}R_x \left[ \frac{75}{16}\frac{R_x^2 + R_y^2}{R^2}
-\frac{15}{4} \right]
\left( \frac{a}{R} \right)^{2},
\label{eq:ave1}\\
\left< \frac{dk}{dt} \right> &\simeq& \frac{GM_2}{R^3}
\sqrt{\frac{a}{GM_1}}R_y \left[ \frac{75}{16}\frac{R_x^2 + R_y^2}{R^2}
-\frac{15}{4} \right]
\left( \frac{a}{R} \right)^{2},
\label{eq:ave2}\\
\left< \frac{dp}{dt} \right> &\simeq& \frac{GM_2}{R^3}
\sqrt{\frac{a}{GM_1}}R_y \left[ \frac{3}{2}\frac{R_z}{R} \right]
\left( \frac{a}{R} \right),
\label{eq:ave3}\\
\left< \frac{dq}{dt} \right> &\simeq& \frac{GM_2}{R^3}
\sqrt{\frac{a}{GM_1}}R_x \left[ \frac{3}{2}\frac{R_z}{R} \right]
\left( \frac{a}{R} \right).
\label{eq:ave4}
\end{eqnarray}

Finally, we integrate them along the trajectory of the passing star: 
\begin{equation}
\left( \begin{array}{c}
	R_x \\R_y \\ R_z 
       \end{array} \right)
=
\left( \begin{array}{c}
	R \cos( f_* + \omega_*)\\
	R \sin( f_* + \omega_*) \cos i_*\\
	R \sin( f_* + \omega_*) \sin i_*
       \end{array} \right), 
\end{equation}
where $f_*$ is the true anomaly of the passing star. 
We integrate Eqs. (\ref{eq:eq_gauss1}) to (\ref{eq:eq_gauss4}) 
over time $t$ from $- \infty$ to $\infty$.  

We integrate $\langle dh/dt \rangle $, $\langle dk/dt \rangle$, 
$\langle dp/dt \rangle$ and $\langle dq/dt \rangle$ 
by $f_*$ instead of by $t$, 
using $df_*/dt$ $=$ $\sqrt{G(M_1+M_2)(e_*+1)D}/R^2$ 
and $R$ $=$ $(e_*+1)D/(1+e_* \cos f_*)$, 
\begin{eqnarray}
\Delta h &\simeq& \int^{\infty}_{-\infty} \left< \frac{dh}{dt} \right> dt 
 = \int_{-\gamma}^{\gamma} \left< \frac{dh}{dt} \right> 
\frac{R^2}{\sqrt{G(M_1+M_2)(e_*+1)D}} df_*
\nonumber\\
&=& - \frac{15}{4 (e_*+1)^{5/2}} \frac{M_*}{\sqrt{M_*+1}} 
\left(\frac{a}{D}\right)^{5/2} \cos \omega_* 
\left[ \frac{1}{8} \gamma e_*^4 (5 \cos^2 i_* -1)  + f  \right],
\label{eq:app_h_e}\\
\Delta k &\simeq& \frac{15}{4 (e_*+1)^{5/2}} \frac{M_*}{\sqrt{M_*+1}} 
\left(\frac{a}{D}\right)^{5/2} \sin \omega_* \cos i_* 
 \left[ \frac{1}{8} \gamma e_*^4 (15 \cos^2 i_* -11)  
+ g  \right],
\\
\Delta p &\simeq& \frac{3}{8 (e_*+1)^{3/2}} \frac{M_*}{\sqrt{M_*+1}} 
\left(\frac{a}{D}\right)^{3/2} (2 \gamma + h ) \sin 2 i_*, \\
\Delta q &\simeq& \frac{e_*}{2} \left(\frac{1-e_*^{-2}}{e_*+1}\right)^{5/2} 
\frac{M_*}{\sqrt{M_*+1}} 
\left(\frac{a}{D}\right)^{3/2} \sin i_* \sin 2 \omega_*,
\label{eq:app_q_e}
\end{eqnarray}
where $\gamma$ is $\cos^{-1}(-1/e_*)$ and  
\begin{eqnarray}
f &=& \sqrt{e_*^{2}-1} \left[ 
\frac{e_*^2}{6} (1+2 e_*^2) 
- \frac{1}{24}(2+e_*^2+12e_*^4) \sin^2 i_*
+ \frac{1}{6}(e_*^2-1)^2 \cos 2 \omega_* \sin^2 i_* \right],\\
g &=& \sqrt{e_*^{2}-1} \left[
\frac{e_*^2}{6} (1+2 e_*^2) 
+ \frac{1}{24}(2 - 19 e_*^2 - 28 e_*^4) \sin^2 i_*
+ \frac{1}{6}(e_*^2-1)^2 \cos 2 \omega_* \sin^2 i_* \right],\\
h &=& \frac{2}{3} e_* \sqrt{1-e_*^{-2}} \{ 
(e_*^{-2} -1) \cos 2 \omega_* + 3 \}.
\end{eqnarray}

Since we start with $h,k,p,q = 0$, the changes are equal to final $h$, $k$,
 $p$ and $q$ of the planetesimals. We obtain final $e$ and $i$ with 
$e = \sqrt{ h^2 + k^2}$ and $i = \sqrt{p^2 + q^2}$. 
For $e_* =1$, $f=g=h=\Delta q = 0$, $e$ and $i$ are 
 \begin{eqnarray}
 e &\simeq& \frac{15 \pi}{128 \sqrt{2}}
\frac{M_*}{\sqrt{1+M_*}}
 \left( \frac{a}{D} \right)^{\frac{5}{2}}
 \nonumber
 \\ && \left[ \cos^2 \omega_*
 \left\{ 5 \cos^2 i_* - 1 \right\}^2  + \sin^2 \omega_* \cos^2 i_*
 \left\{  15 \cos^2 i_* - 11 \right\}^2
 \right]^{\frac{1}{2}},
\label{eq:e1_app} 
 \\
 i &\simeq& \frac{3 \alpha}{8 \sqrt{2}} \frac{M_*}{\sqrt{1+M_*}}
 \left( \frac{a}{D} \right)^{\frac{3}{2}} \sin 2 i_*.
 \label{eq:i1_app}
 \end{eqnarray}

\section*{ACKNOWLEDGMENTS}

We thank Kiyoshi Nakazawa and Hiroyuki Emori for useful scientific discussions. 
We also thank John D. Larwood and Giovanni B. Valsecchi for critical comments, 
which helped us improve the original version of the manuscript, and 
Andeas Burkert for motivating us to study this problem. 
This work is supported in part by Grant-in-Aid for Specially Promoted Research 
of the Japanese Ministry of Education, Science, and Culture (11101002).

\newpage
\figcaption{
Encounter configuration in the frame centered at the primary star with mass $M_1$. 
The orbit of the passing star with mass $M_2$ 
is characterized by pericenter distance $D$, eccentricity $e_*$, 
inclination $i_*$
 and argument of perihelion 
$\omega_*$. 
If length and mass are scaled by $D$ and $M_1$, the encounter parameter are 
$M_*$ ( = $M_2/M_1$), $e_*$, $i_*$ and $\omega_*$. 
\label{config}
}

\figcaption{Left and center panels are the time evolution of orbital 
eccentricity $e$ and inclination $i$ of particles as a function of 
scaled initial semimajor axis $a/D$. 
Time proceeds from (a) top to (c) bottom panels. 
The right panels are face-on snapshots of the disk particles 
(small dots) 
and the passing star (filled circle).
\label{time_d}
}
\figcaption{
(a)Orbital eccentricity $e$ and inclination $i$ of particles pumped-up 
by a passing star, as a function of initial scaled semimajor axis $a/D$, 
in the case with $i_* = 5^{\circ}$, $\omega_* = 90^{\circ}$, 
$e_* = 1$ and $M_*=1$. (b)The results with $e_* = 1$ and $M_*= 0.2$ 
($i_*$ and $\omega_*$ are the same).
(c)The result with $e_* = 5$ and $M_*=1$. 
Dashed lines with triangles express analytical expression 
given by Eqs. (\ref{eq:app_h_e}) to (\ref{eq:app_q_e}) in Appendix. 
The dashed lines (in particular, for $i$) are almost indistinguishable from 
the numerical results in the inner region. 
\label{M_e_dep}
}

\figcaption{
Orbital eccentricity $e$ and inclination $i$ of particles pumped-up 
by a passing star, as a function of scaled semimajor axis $a/D$, 
in the case with  $\omega_* = 0^{\circ}$, $e_* = 1$ and $M_*=1$.
Orbital inclination of the passing star is (a) $5^{\circ}$, 
(b) $30^{\circ}$, (c) $45^{\circ}$, (d) $85^{\circ}$ and (e) $150^{\circ}$. 
Dashed lines with triangle express analytical expression given by 
Eqs. (\ref{eq:e}) 
and (\ref{eq:i}) in Section 5. 
The dashed lines are almost distinguishable from the numerical results. 
\label{is_dep}
}

\figcaption{
Numerically and analytically calculated $\tilde{e}_0$ 
are plotted, $\tilde{e}_0$ is defined 
by $e$ $=$ $(15 \pi / 64)$ $\tilde{e}_0$ $(a/D)^{5/2}$ in the inner disk, 
in the case with $e_* = 1$ and $M_* = 1$. 
(a)The contours of $\tilde{e}_0$ numerical calculated 
at every 10 degrees of $i_*$ and $\omega_*$. 
(b)The contours of analytically calculated $\tilde{e}_0$: $[\cos^2 \omega_*
 \{ 1 - (5/4) \sin^2 i_*  \}^2$ $ +$ $\sin^2 \omega_* \cos^2 i_*
 \{  1 - (15/4) \sin^2 i_* \}^2]^{1/2}$ (Eq. (\ref{eq:e})), as a function of 
$i_*$ and $\omega_*$. 
\label{e0_cont}
}

\figcaption{
Numerically and analytically calculated $\tilde{i_0}$ are plotted, 
where $\tilde{i}_0$ is defined by 
$i$ $=$ $3 \pi / 16$ $\tilde{i}_0$ $(a/D)^{3/2}$ in the inner disk, 
in the case with $e_* = 1$ and $M_* = 1$. 
Circles are numerical results. 
We calculated the case with $\omega_*$ at every $10^{\circ}$ 
from $0^{\circ}$ to $90^{\circ}$, 
for each $i_*$. 
The results overlap each other: 
$\tilde{i}_0$ is almost completely independent of $\omega_*$. 
Dashed lines express analytical expression of $\tilde{i}_0$: $ \sin 2 i_*$ 
(see Eq. (\ref{eq:i})). 
\label{i0_cont}
}

\figcaption{The relative change in semimajor axis $\Delta a$ 
after a stellar encounter 
as a function of initial semimajor axis 
$a$ in the case with $i_* =5^{\circ}$, 
$\omega_* = 90^{\circ}$, $M_*=1$ and $e_*=1$.
\label{a0_vs_a}
}

\figcaption{
(a) orbital eccentricity $e$, (b) inclination $i$, 
(c) longitude of pericentre $\tilde{\omega}$ and (d) longitude of 
ascending node $\varphi$ 
of planetesimals, as a function of initial semimajor axis $a/D$, 
in the case with initial $e$ and $i$,
$i_* = 5^{\circ}$, $\omega_* = 0^{\circ}$, 
$e_* = 1$ and $M_*=1$.
\label{e0_omega}
}

\figcaption{
(a)$e\sub{eff}$ and $i\sub{eff}$, 
(b)$e\sub{rel}$ and $i\sub{rel}$ and 
(c)$e$ and $i$, 
of the colliding planetesimals, 
as a function of initial semimajor axis $a/D$, where we define $a$ 
is the radius of the colliding point. 
($e_0$, $i_0$) = (0, 0) and $(i_*, \omega_*, M_*, e_*) = (5^{\circ}, 
 0^{\circ}, 1, 1)$.
\label{e0_rel}
}

\figcaption{
The same as Fig. \ref{e0_omega} except for 
$e_0$ and $i_0$ are $1 \times 10^{-4}$.
\label{e5_omega}
}

\figcaption{
The same as Fig. \ref{e0_rel} except for 
$e_0$ and $i_0$ are $1 \times 10^{-4}$.
\label{e5_rel}
}

\begin{figure}
\plotone{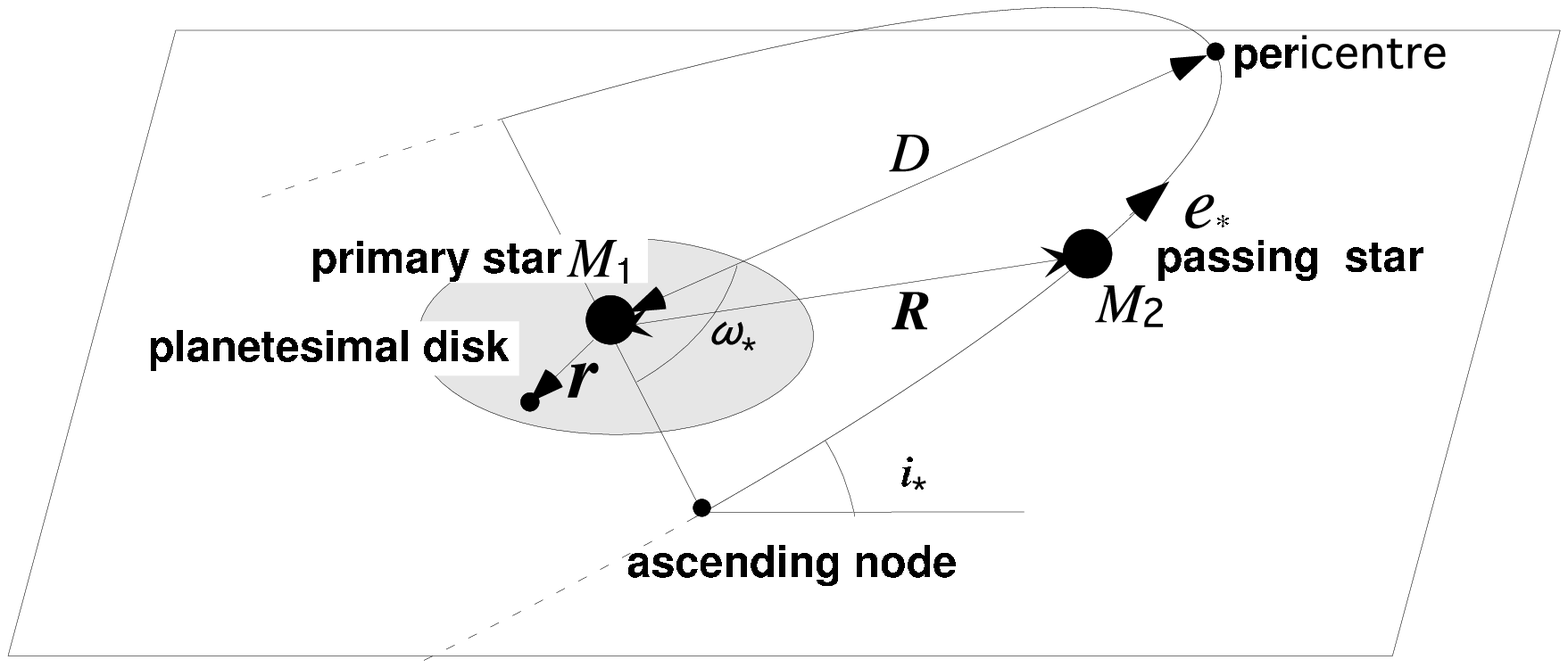}
Fig. \ref{config} --- Kobayashi and Ida (2001)
\end{figure}

\begin{figure}
\plotone{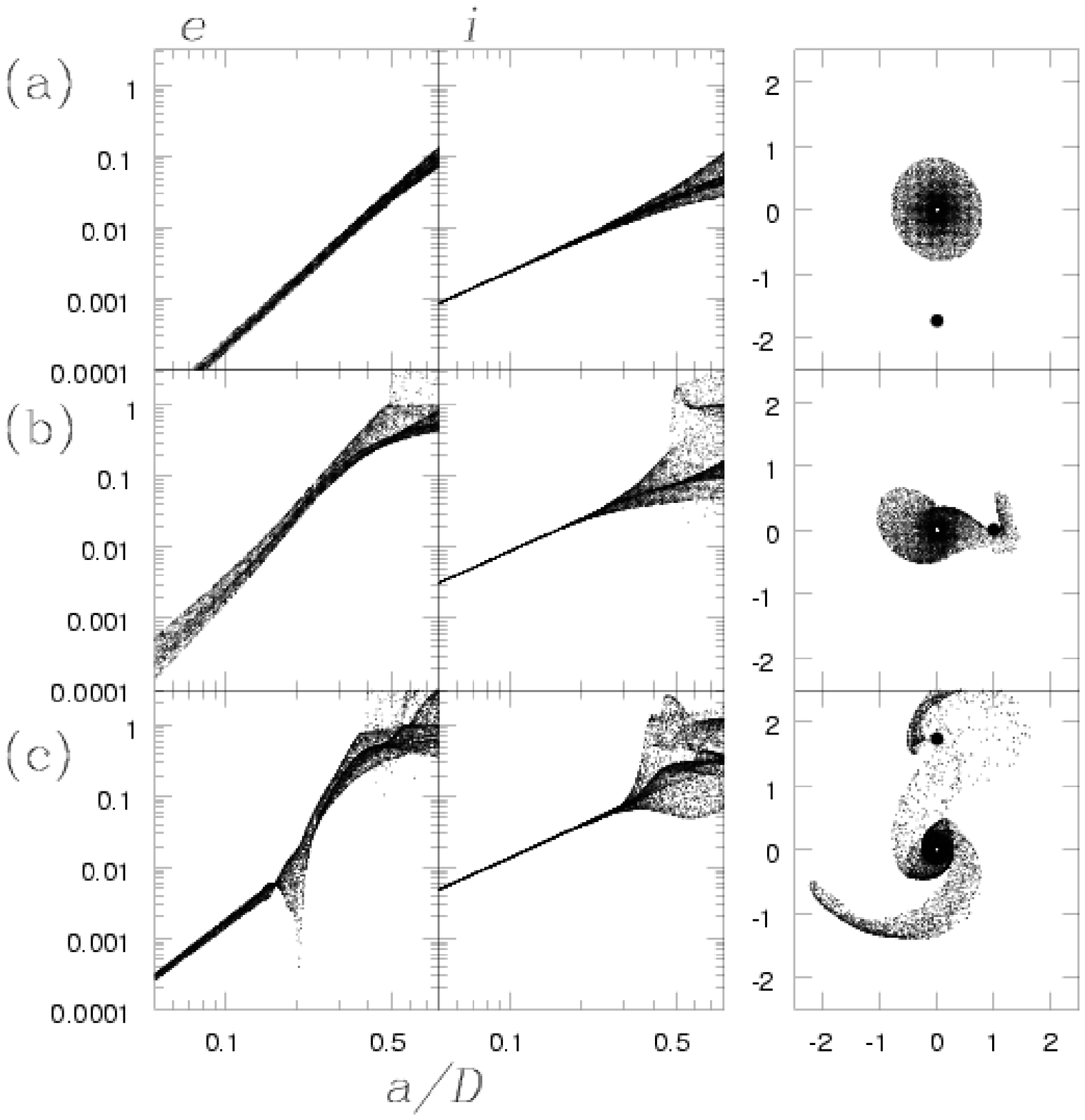}
Fig. \ref{time_d} --- Kobayashi and Ida (2001)
\end{figure}

\begin{figure}
\plotone{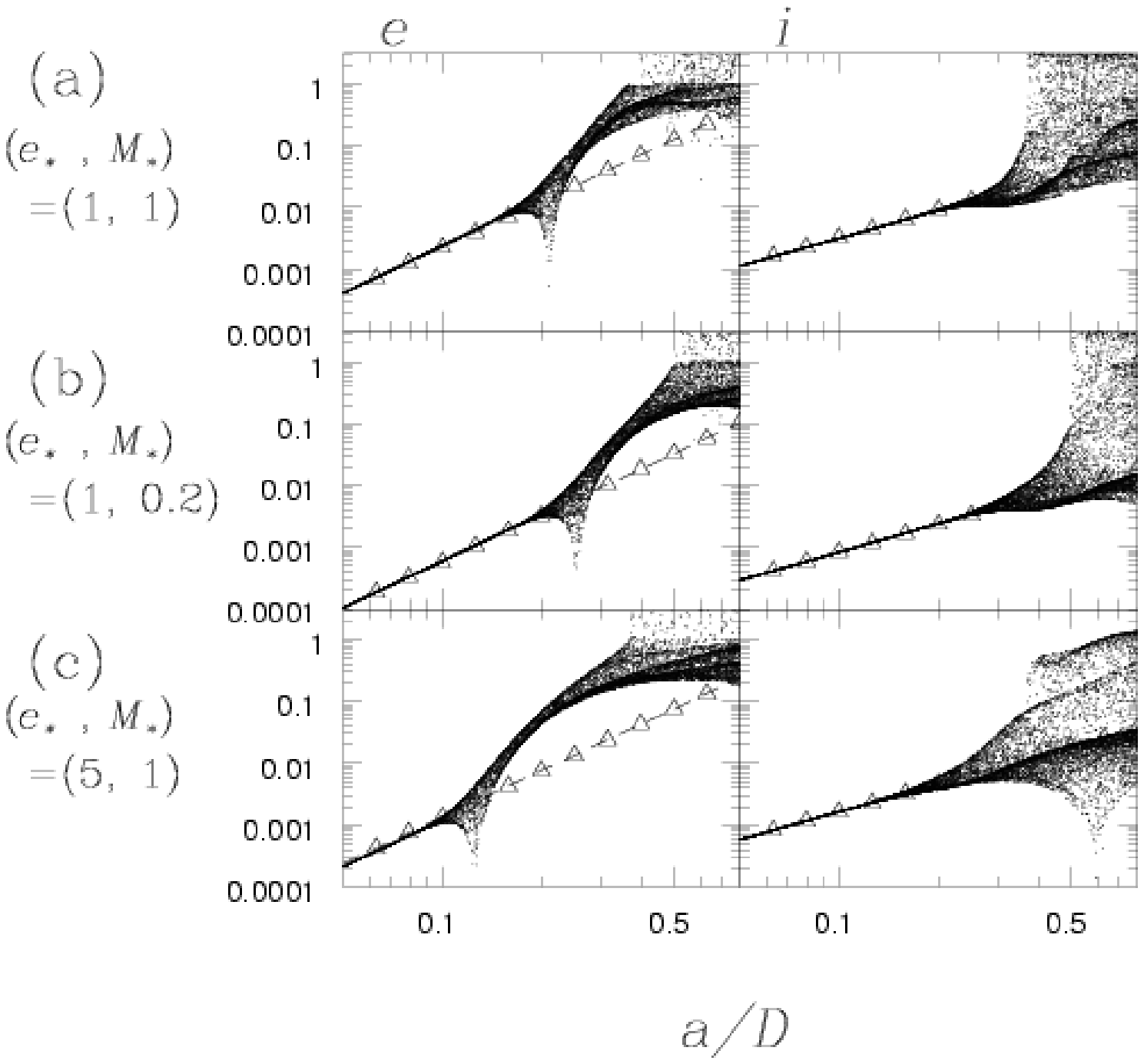}
Fig. \ref{M_e_dep} --- Kobayashi and Ida (2001)
\end{figure}

\begin{figure}
\plotone{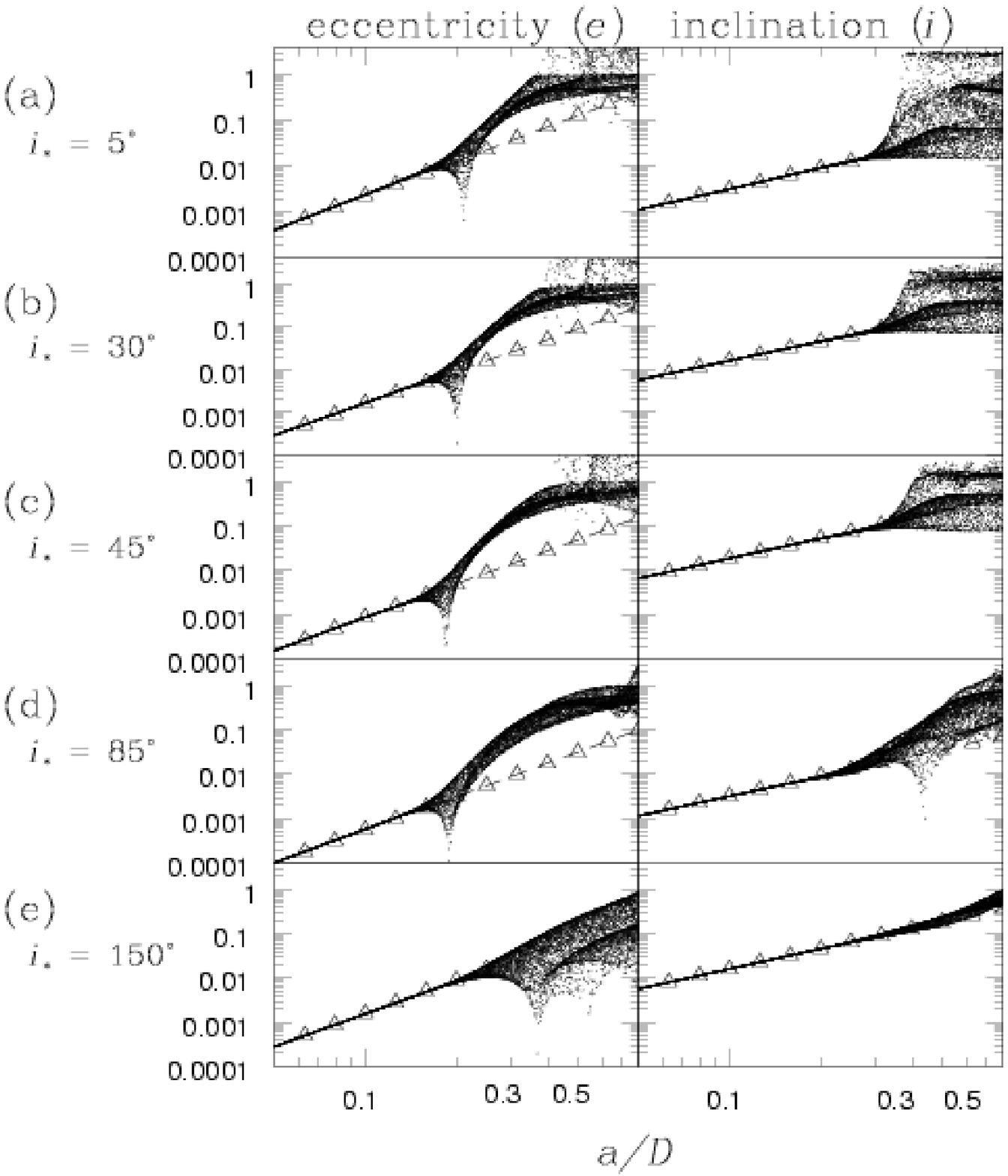}
Fig. \ref{is_dep} --- Kobayashi and Ida (2001)
\end{figure}

\begin{figure}
\plotone{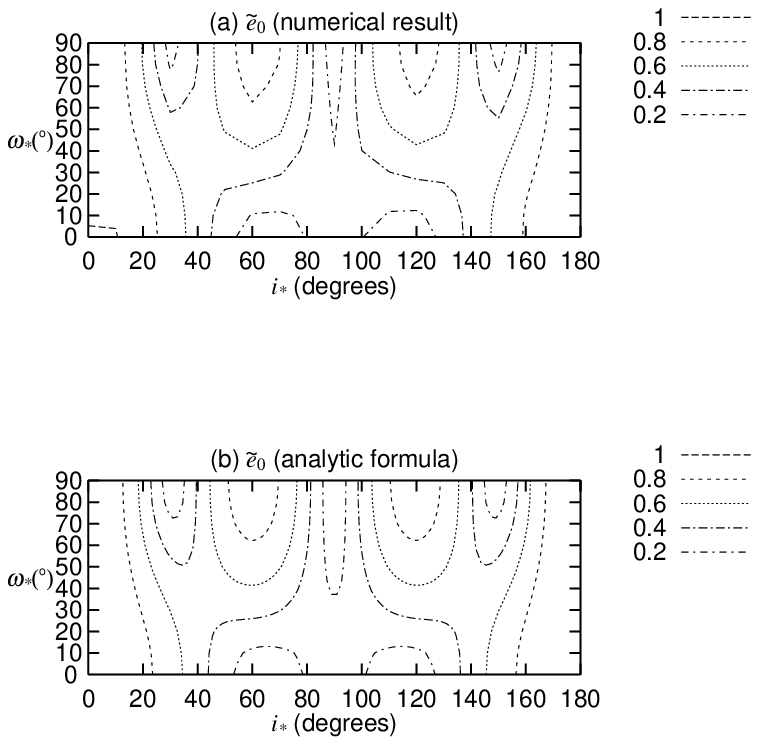}
Fig. \ref{e0_cont} --- Kobayashi and Ida (2001)
\end{figure}

\begin{figure}
\plotone{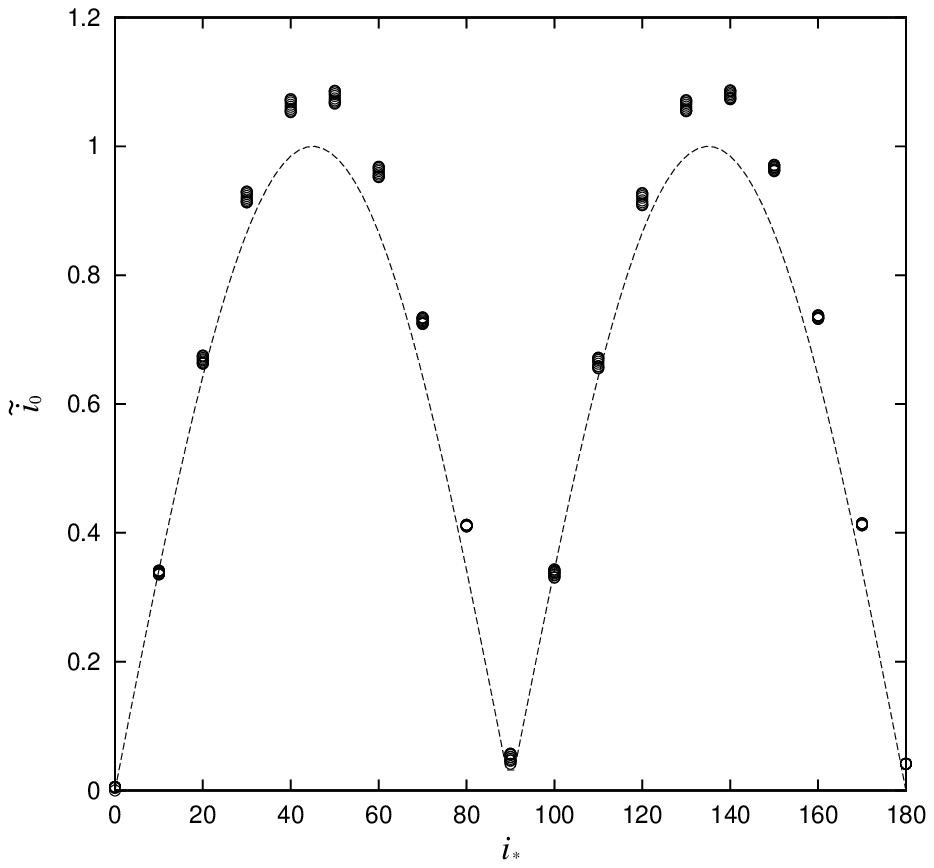}
Fig. \ref{i0_cont} --- Kobayashi and Ida (2001)
\end{figure}

\begin{figure}
\epsscale{1}
\plotone{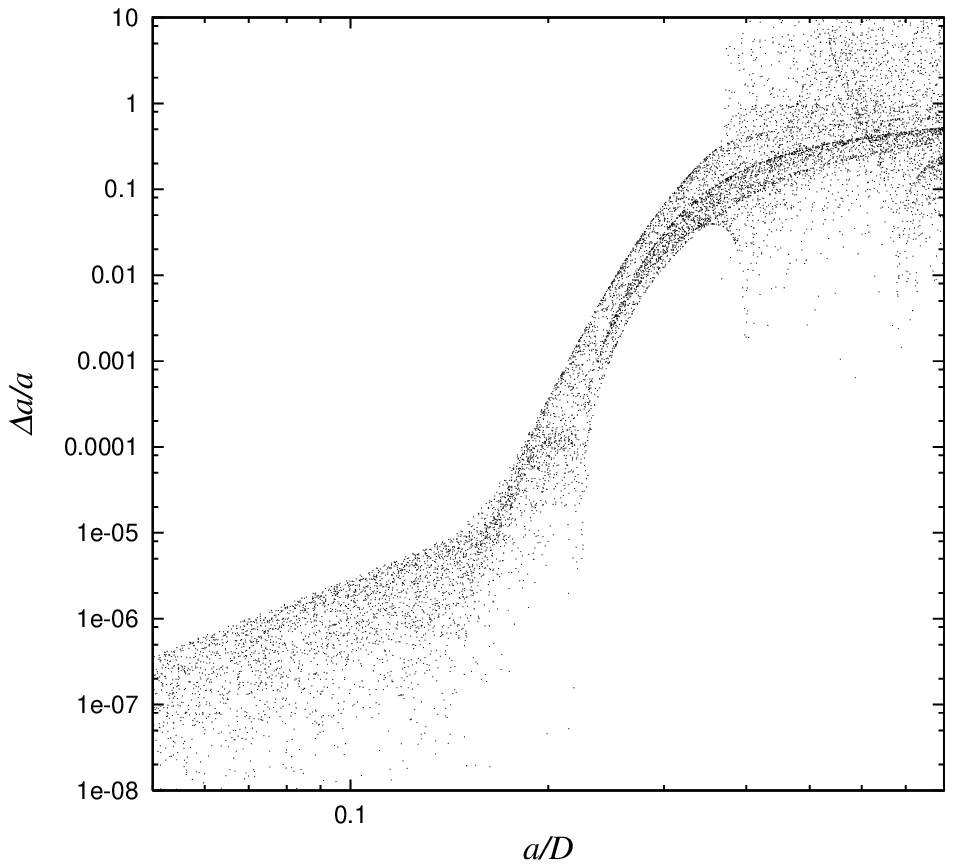}
Fig. \ref{a0_vs_a} --- Kobayashi and Ida (2001) 
\end{figure}

\begin{figure}
\plotone{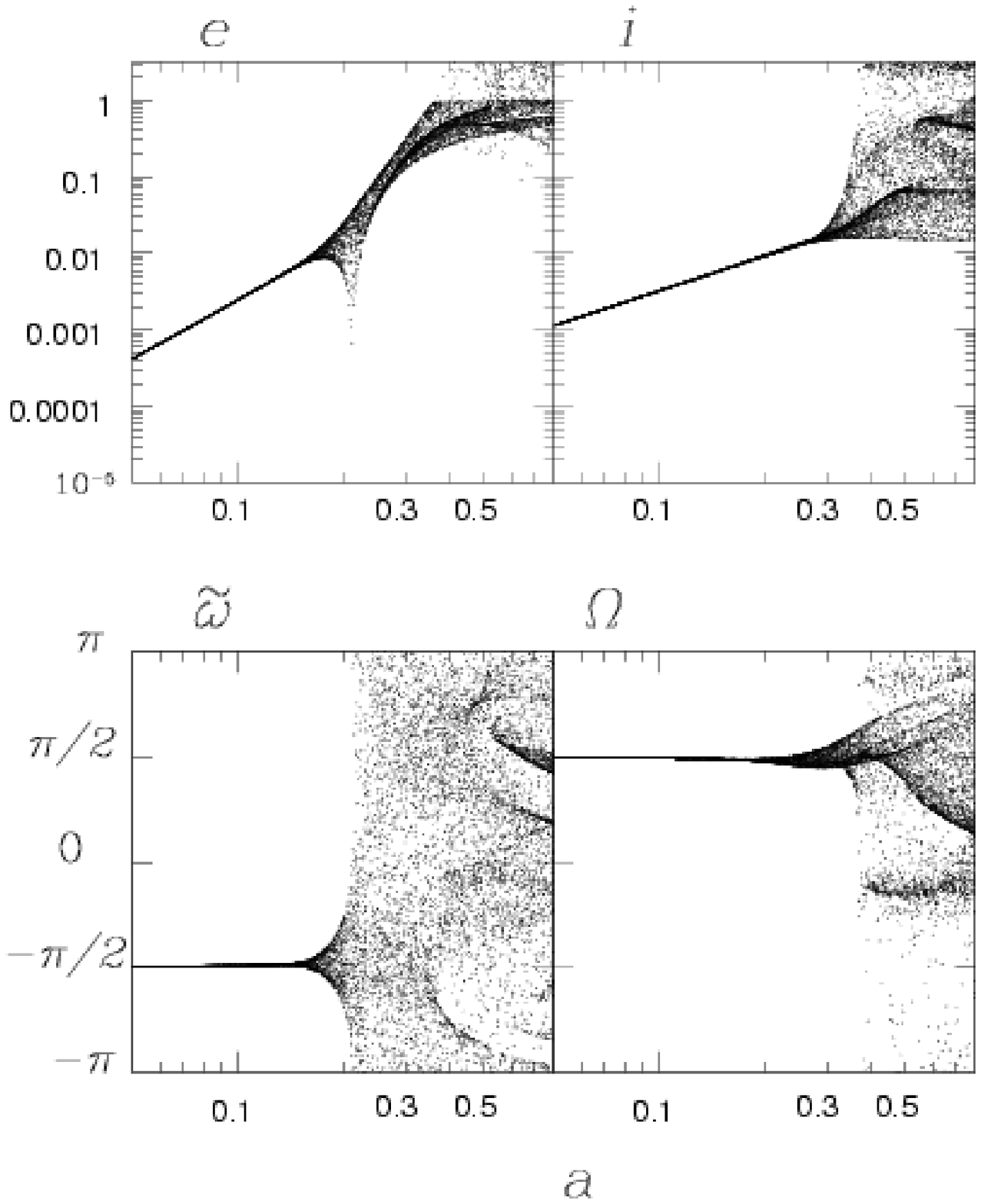}
Fig. \ref{e0_omega} --- Kobayashi and Ida (2001) 
\end{figure}

\begin{figure}
\plotone{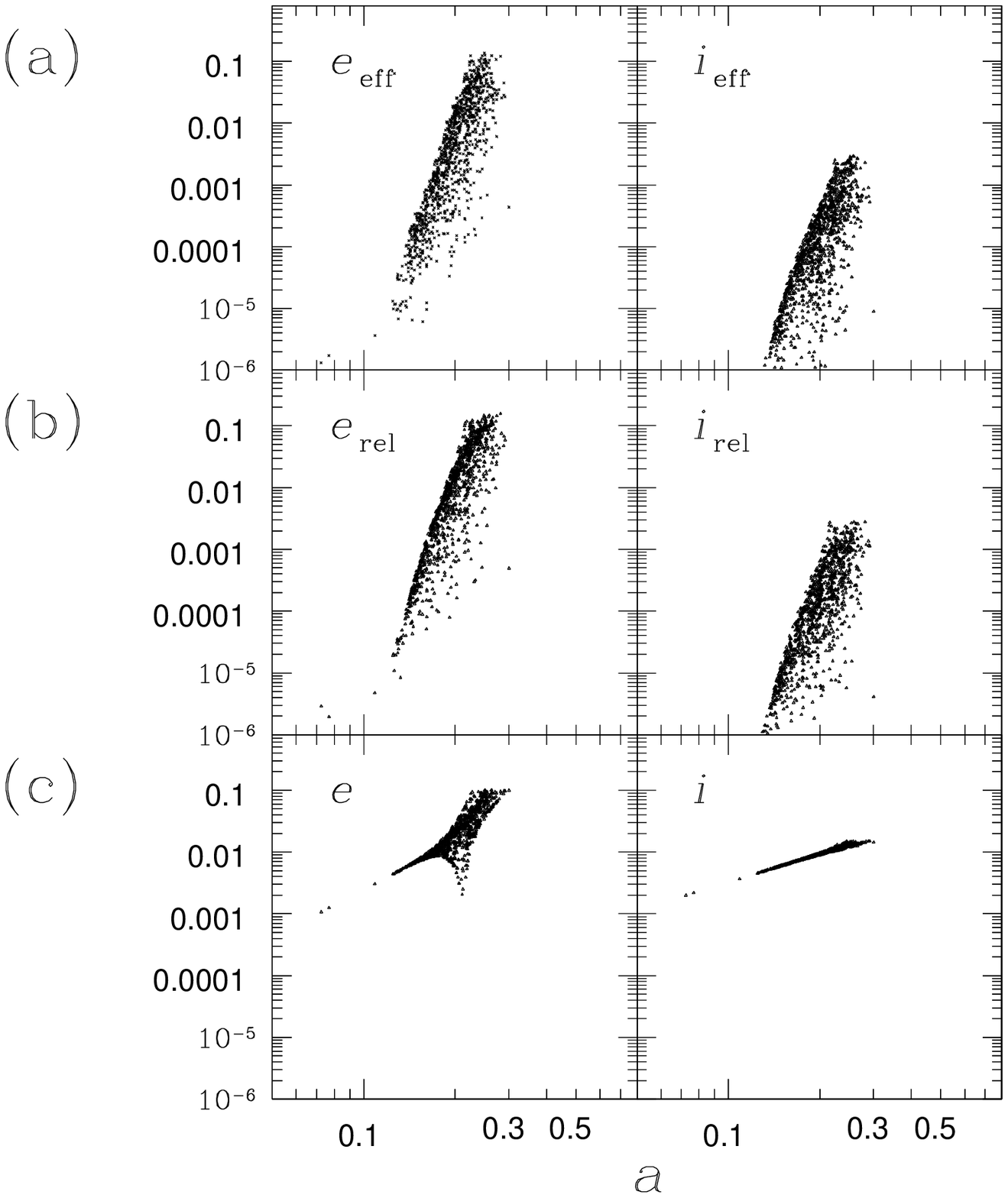}
Fig. \ref{e0_rel} --- Kobayashi and Ida (2001) 
\end{figure}

\begin{figure}
\plotone{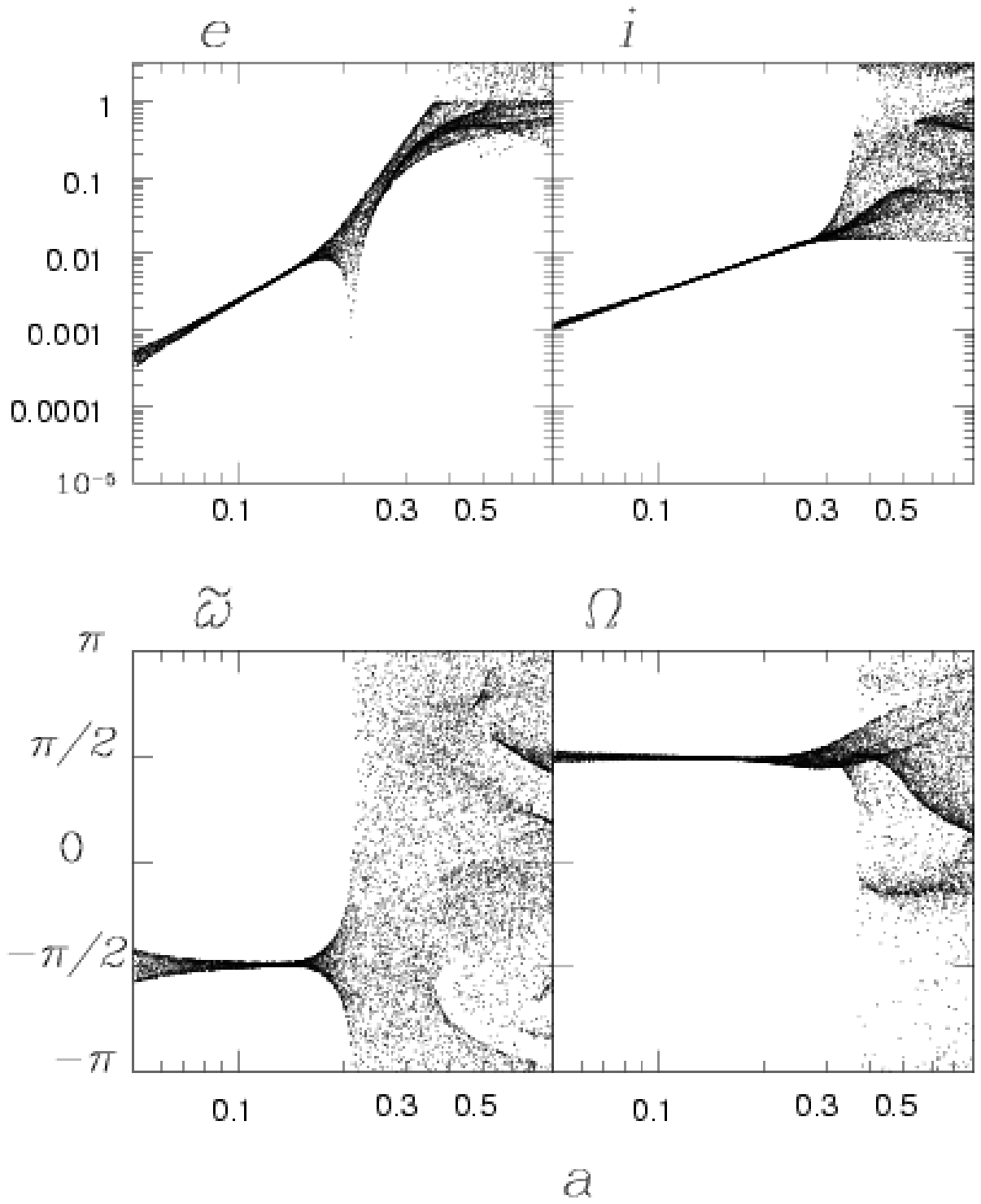}
Fig. \ref{e5_omega} --- Kobayashi and Ida (2001) 
\end{figure}

\begin{figure}
\plotone{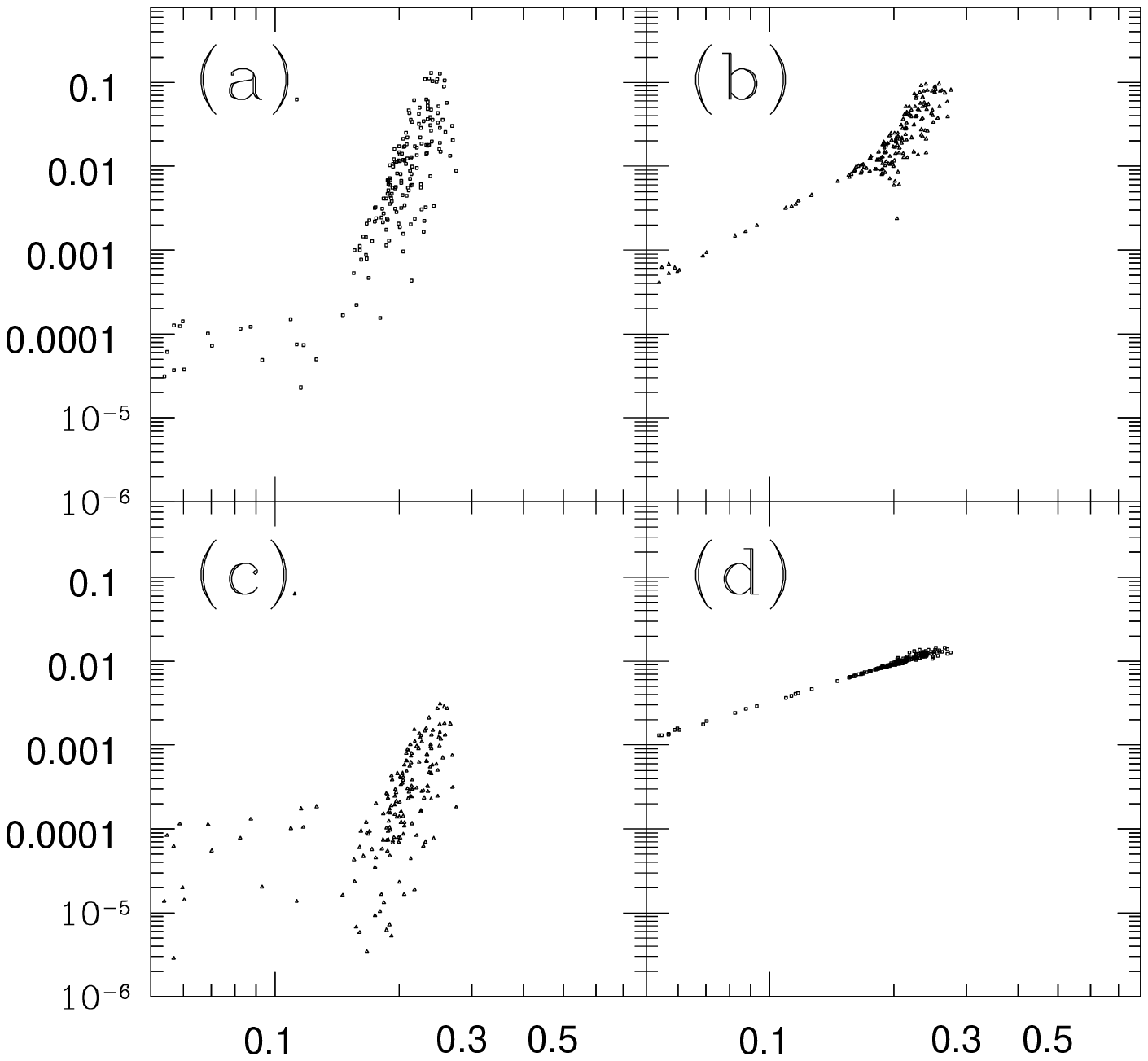}
Fig. \ref{e5_rel} --- Kobayashi and Ida (2001) 
\end{figure}


\begin{thebibliography}{}
\bibitem[Adachi {\it et al.}(1976)]{adachi}
Adachi, I., Hayashi, C., and Nakazawa, K. 1976. 
The gas drag effect on the elliptical motion of a solid body in the 
primordial solar nebula.
{\it Prog. Theor. Phys.} {\bf 56} 1756-1771.
\bibitem[Adams and Laughlin(2001)]{adams00}
Adams, F. C., and G. Laughlin 2001. Constraints on the Birth Aggregate of the Solar System. {\it Icarus} {\bf 150} 151-162 
\bibitem[Andrillat {\it et al.}(1990)]{andrillat90}
Andrillat, Y., M. Jaschek, and  C. Jaschek 1990. The infrared spectrum of HD 141569. {\it Astron. Astrophys.} {\bf 233}, 474-476.
\bibitem[Artymowicz(1997)]{artymo97}
Artymowicz, P. 1997. Beta Pictoris: an Early Solar System? {\it Annu. Rev. Earth Planet. Sci.} {\bf 25}, 175-219. 
\bibitem[Augereau {\it et al.}(1999)]{aug99}
Augereau, J. C., A. M. Lagrange, D. Mouillet, and F. Mnard 1999. 
ST/NICMOS2 observations of the HD 141569 A circumstellar disk.  
{\it Astron. Astrophys.} {\bf 350}, L51-L54.
\bibitem[Backman and Paresce(1993)]{backman95}
Backman, D. E., and F. Paresce 1993.
Main-sequence stars with circumstellar solid material: the Vega phenomenon. 
In {\it Protostars and planets III} (E. H. Levy and J. Lunine, Eds.), 
pp. 1253-1304. Univ. of Arizona Press, Tucson.
 {\it Astrophys. J.} {\bf 450}, 1253-1304.
\bibitem[Barnes and Hernquist(1992)]{barn92}
Barnes, J. E., and  L. Hernquist 1992. 
Dynamics of interacting galaxies. {\it Annu. Rev. Astron. Astrophys.} {\bf 30}, 705-742. 
\bibitem[Beckwith and Sargent(1996)]{beckwith1996}
Beckwith,S. V. W., and A. I. Sargent 1996. 
Circumstellar disks and the search for neighbouring planetary systems. 
{\it Nature} {\bf 383}, 139-144. 
\bibitem[Binney and Tremaine(1987)]{binney1987}
Binney, J., and S. Tremaine 1987. {\it Galactic Dynamics.} 
Princeton Univ. Press, Princeton.
\bibitem[Boffin {\it et al.}(1998)]{boffin98}
Boffin, H. M. J., S. J. Watkins, A. S. Bhattal, N. Francis, and A. P. Whitworth, 
1998. 
Numerical simulations of protostellar encounters - I. Star-disc encounters. 
{\it Mon. Not. R. Astron. Soc.} {\bf 300}, 1189-1204. 
\bibitem[Bodenheimer and Pollack(1986)]{bodenheimer86}
Bodenheimer, P., and J. B. Pollack 1986.
Calculations of the accretion and evolution of giant planets The effects of solid cores. 
{\it Icarus}, {\bf 67}, 391-408. 
\bibitem[Bonnell {\it et al.}(2001)]{bonnell01}
Planetary dynamics in stellar clusters. 
{\it Mon. Not. R. Astron. Soc.} {\bf 322}, 859-865.
\bibitem[Brouwer and Clemence(1961)]{brouwer}
Brouwer, D., and G. M. Clemence, 1961. {\it Methods of Celestial Mechanics.} 
Academic Press, New York.
\bibitem[Brunini and Fern\'{a}ndez(1996)]{brunini96}
Brunini, A., and A. Fern\'{a}ndez, 1996. 
Perturbations on an extended Kuiper disk caused by passing stars and giant molecular clouds. 
{\it Astron. Astrophys.} {\bf 308}, 988-994. 
\bibitem[Clarke and Pringle(1991)]{clarke91}
Clarke, C. J., and J. E. Pringle 1991.
Star-disc interactions and binary star formation.
{\it Mon. Not. R. Astron. Soc.} {\bf 249}, 584-587.
\bibitem[Clarke and Pringle(1993)]{clarke93}
Clarke, C. J., and J. E. Pringle, 1993. 
Accretion disc response to a stellar fly-by. 
{\it Mon. Not. R. Astron. Soc.} {\bf 261}, 190-202. 
\bibitem[de la Fuente Marcos and de la Fuente Marcos(1997)]{marcos97}
de la Fuente Marcos, C. and R. de la Fuente Marcos, 1997. 
Eccentric giant planets in open star clusters. 
{\it Astron. Astrophys.} {\bf 326}, L21-L24. 
\bibitem[Eggers and Woolfson(1996)]{eggers96}
Eggers, S., \& M. M. Woolfson 1996.  
Stellar perturbations of inner core comets and the impulse approximation.
{\it Mon. Not. R. Astron. Soc.} {\bf 282}, 13-18.  
\bibitem[Fajardo-Acosta {\it et al.}(1998)]{faj98}
Fajardo-Acosta, S. B., C. M. Telesco, C. M., and R. F. Knacke 1998. 
Infrared Photometry of beta Pictoris Type Systems. 
{\it Astron. J.} {\bf 115}, 2101-2121.  
\bibitem[Fern\'{a}ndez(1997)]{fern97}
Fern\'{a}ndez, A. J. 1997, 
The Formation of the Oort Cloud and the Primitive Galactic Environment.
{\it Icarus} {\bf 129}, 106-119. 
\bibitem[Goldreich and Tremaine(1982)]{gt82}
Goldreich, P., and S. Tremaine 1982. 
The dynamics of planetary rings. 
{\it Annu. Rev. Astron. Astrophys.} {\bf 20} 249-283.
\bibitem[Goldreich and Ward(1973)]{goldreich73}
Goldreich, P., and W. R. Ward 1973. The Formation of Planetesimals. 
{\it Astrophys. J.} {\bf 183}, 1051-1062    
\bibitem[Greaves {\it et al.}(1998)]{greav98}
Greaves, J. S., and 10 colleagues, 1998. 
A Dust Ring around epsilon Eridani: Analog to the Young Solar System. 
{\it Astrophys. J.} {\bf 506}, L133-L137. 
\bibitem[Greenberg {\it et al.}(1978)]{greenberg78}
Greenberg, R., W. K. Hartmann, C. R. Chapman, and J. F. Wacker, 1978. 
Planetesimals to planets - Numerical simulation of collisional evolution. 
{\it Icarus} {\bf 35}, 1-26. 
\bibitem[Hall(1997)]{hall97}
Hall, S. M. 1997. 
Circumstellar disc density profiles: a dynamic approach. 
{\it Mon. Not. R. Astron. Soc.} {\bf 287}, 148-154. 
\bibitem[Hall {\it et al.}(1996)]{hall96}
Hall, S. M., C. J. Clarke, and J. E. Pringle 1996. 
Energetics of star-disc encounters in the non-linear regime. 
{\it Mon. Not. R. Astron. Soc.} {\bf 278}, 303-320. 
\bibitem[Hayashi(1981)]{hayashi81}
Hayashi, C. 1981.
Structure of the solar nebula, growth and decay of magnetic fields and effects of magnetic and turbulent viscosities on the nebula. 
{\it Prog. Theor. Phys. Suppl.} {\bf 70}, 35-53. 
\bibitem[Hayashi {\it et al.}(1985)]{hayashi85} 
Hayashi, C., K. Nakazawa, and Y. Nakagawa 1985. 
Formation of the solar system. 
In {\it Protostars and Planets II} (D. C. Black and M. S. Matthews, Eds.), 
pp. 1100-1153. Univ. of Arizona Press, Tucson. 
\bibitem[Heggie and Rasio(1996)]{heggie96}
Heggie, D. C., and F. A. Rasio 1996. 
The Effect of Encounters on the Eccentricity of Binaries in Clusters.
{\it Mon. Not. R. Astron. Soc.}, {\bf 282}, 1064-1084.
\bibitem[Heller(1993)]{heller93}
Heller, C. H. 1993. 
Encounters with protostellar disks. I - Disk tilt and the nonzero solar obliquity. {\it Astrophys. J.} {\bf 408}, 337-346.
\bibitem[Heller(1995)]{heller95}
Heller, C. H. 1995.
Encounters with Protostellar Disks. II. Disruption and Binary Formation. 
{\it Astrophys. J.} {\bf 455}, 252-259
\bibitem[Heppenheimer(1978)]{heppen}
Heppenheimer, T. A. 1978. 
On the formation of planets in binary star systems.
{\it Astron. Astrophys.} {\bf 65}, 421-426. 
\bibitem[Ida {\it et al.}(2000)]{ida00}
Ida, S., J. Larwood, and A. Burkert 2000. 
Evidence for Early Stellar Encounters in the Orbital Distribution of Edgeworth-Kuiper Belt Objects. 
{\it Astrophys. J.} {\bf 528}, 351-356. 
\bibitem[Ikoma {\it et al.}(2000)]{ikoma00}
Ikoma, M., K. Nakazawa, and H. Emori 2000. 
Formation of Giant Planets: Dependences on Core Accretion Rate and Grain Opacity.
{\it Astrophys. J.} {\bf 537}, 1013-1025. 
\bibitem[Ito and Tanikawa(2001)]{ito01}
Ito, T., and Tanikawa, K. 2001.
Stability of Terrestrial Protoplanet Systems and Alignment of Orbital Elements.
{\it PASJ} {\bf 53}, 143-151. 
\bibitem[Jayawardhana {\it et al.}(1998)]{jaya98}
Jayawardhana, R., S. Fisher, L. Hartmann, C. Telesco, R. Pina, and G. Fazio 1998. 
A Dust Disk Surrounding the Young A Star HR 4796A. 
{\it Astrophys. J.} {\bf 503}, L79-L82. 
\bibitem[Jura {\it et al.}(1995)]{jura95}
Jura, M., A. M. Ghez,R. J.  White, D. W. McCarthy, R. C. Smith, and P. G. Martin 
1995. The fate of the solid matter orbiting HR 4796A. 
{\it Astrophys. J.} {\bf 455}, 451-456. 
\bibitem[Jourdain de Muizon {\it et al.}(1999)]{jour99}
Jourdain de Muizon, M., and 10 colleagues, 1999. 
A very cold disc of dust around the G0V star HD 207129. 
{\it Astron. Astrophys.} {\bf 350}, 875-882. 
\bibitem[Kalas {\it et al.}(2000)]{kalas00}
Kalas, P., J. Larwood, B. A. Smith, and A. Schultz 2000. 
Rings in the Planetesimal Disk of $\beta$ Pictoris.
{\it Astrophys. J.} {\bf 530}, 133-137. 
\bibitem[Kalas and Jewitt(1995)]{kalas95}
Kalas, P., and D. Jewitt 1995. 
Asymmetries in the Beta Pictoris Dust Disk
{\it Astron. J.} {\bf 110}, 794-804.  
\bibitem[Korycansky and Papaloizou(1995)]{kory95}
Korycansky, D. G., and J. C. B. Papaloizou 1995. 
The response of a gaseous disc to a binary encounter
{\it Mon. Not. R. Astron. Soc.} {\bf 274}, 85-98. 
\bibitem[Kroupa(1995)]{kroupa1995}
Kroupa, P. 1995. 
The dynamical properties of stellar systems in the Galactic disc. 
{\it Mon. Not. R. Astron. Soc.} {\bf 277} 1507-1521.  
\bibitem[Kroupa(1998)]{kroupa1998}
Kroupa, P. 1998. 
On the binary properties and the spatial and kinematical distribution of young stars. 
{\it Mon. Not. R. Astron. Soc.} {\bf 298}, 231-242. 
\bibitem[Larwood and Kalas(2001)]{john00}
Larwood, J. D., and P. G. Kalas 2001. 
Close stellar encounters with planetesimal disks: The dynamics of asymmetry 
in the $\beta$ Pictoris system. 
{\it Mon. Not. R. Astron. Soc.} {\bf 323}, 402-416. 
\bibitem[Larwood(1997)]{larwood97}
Larwood, J. D. 1997. 
The tidal disruption of protoplanetary accretion discs. 
{\it Mon. Not. R. Astron. Soc.} {\bf 290}, 490-504. 
\bibitem[Laughlin and Adams(1998)]{lau98}
Laughlin G., and F. C. Adams 1998. 
The Modification of Planetary Orbits in Dense Open Clusters. 
{\it Astrophys. J.} {\bf 508}, L171-L174. 
\bibitem[Laughlin and Adams(2000)]{lau00}
Laughlin G., and F. C. Adams 2000. 
The frozen Earth: Binary scattering events and the fate of the Solar. 
{\it Icarus}, {\bf 145}, 614-627. 
\bibitem[Lissauer and Stewart(1993)]{lissauer93}
Lissauer, J. J., and G. R. Stewart, 1993.  
Growth of planets from planetesimals. 
In {\it Protostars and planets III} (E. H. Levy and J. Lunine, Eds.), 
pp. 1061-1088. Univ. of Arizona Press, Tucson.
\bibitem[Marzari and Scholl(2000)]{marzari00}
Marzari, F., and H. Scholl 2000. 
Planetesimal Accretion in Binary Star Systems.
{\it Astrophys. J.}, {\bf 543}, 328-339.
\bibitem[Mizuno(1980)]{mizuno80}
Mizuno, H. 1980. 
Formation of the Giant Planets. 
{\it Prog. Theor. Phys.}, {\bf 64}, 544-557. 
\bibitem[Ohtsuki {\it et al.}(1993)]{ohtsuki93}
Ohtsuki, K., S. Ida, Y. Nakagawa, and K. Nakazawa 1993. 
Planetary accretion in the solar gravitational field. 
In {\it Protostars and Planets III} (E. H. Levy \& J. Lunine Eds.), 
pp. 1089-1107. Univ. Arizona Press, Tucson. 
\bibitem[Ohtsuki(1993)]{ohtsu93}
Ohtsuki, K. 1993.
Capture probability of colliding planetesimals - Dynamical constraints on accretion of planets, satellites, and ring particles. 
{\it Icarus} {\bf 106}, 228-246. 
\bibitem[Ostriker(1994)]{ost94}
Ostriker, E. C. 1994. 
Capture and induced disk accretion in young star encounters. 
{\it Astrophys. J.} {\bf 424}, 292-318. 
\bibitem[Palmer and Papaloizou(1982)]{plamer82}
Palmer, P. L. \& J. Papaloizou, 1982. 
Tidal interactions of disc galaxies. 
{\it Mon. Not. R. Astron. Soc.} {\bf 199}, 869-882. 
\bibitem[Safronov(1969)]{safronov69}
Safronov, V. S. 1969. {\it Evolution of the Protoplanetary Cloud
and Formation of the Earth and Planets.} Nauka Press, Moscow. 
\bibitem[Sahu {\it et al.}(1998)]{sahu98}
Sahu, M. S., J. C. Blades, L. He, D. Hartmann, M. J. Barlow, and
I. A. Crawford 1998. 
Atomic and Molecular Interstellar Absorption Lines toward the High Galactic Latitude Stars HD 141569 and HD 157841 at Ultra-High Resolution. 
{\it Astrophys. J.} {\bf 504}, 522-532. 
\bibitem[Stern(1995)]{starn}
Stern, S. A. 1995. Collisional time scales in the Kuiper disk and their implications. {\it Astrophys. J.} {\bf 110}, 856-868.
\bibitem[Toomre and Toomre(1972)]{tt72}
Toomre, A., and J. Toomre 1972. 
Galactic Bridges and Tails. 
{\it Astrophys. J.} {\bf 178}, 623-666. 
\bibitem[Wahde {\it et al.}(1996)]{wahde96}
Wahde, M., K. J. Donner, and B. Sundelius 1996. 
Dynamical friction in disc galaxies with non-zero velocity dispersion. 
{\it Mon. Not. R. Astron. Soc.} {\bf 281}, 1165-1182. 
\bibitem[Weidenschilling and Cuzzi(1993)]{weidenschilling93}
Weidenschilling, S. J., and J. N. Cuzzi 1993. 
Formation of planetesimals in the solar nebula. 
In {\it Protostars and Planets III} (E. H. Levy and J. I. Lunine Eds.), 
pp. 1031-1060. Univ. of Arizona Press, Tucson.
\bibitem[Weinberger {\it et al.}(1999)]{wei99}
Weinberger, A. J., E. E. Becklin, G. Schneider, B. A. Smith, P. J. Lowrance, 
M. D. Silverstone, B. Zuckerman, and R. J. Terrile 1999. 
The Circumstellar Disk of HD 141569 Imaged with NICMOS.
{\it Astrophys. J.} {\bf 525}, L53-L56. 
\bibitem[Weinberger {\it et al.}(2000)]{wei00}
Weinberger, A. J., R. M. Rich, E. E. Becklin, B. Zuckerman, and K. Matthews 
2000. 
Stellar Companions and the Age of HD 141569 and Its Circumstellar Disk. 
{\it Astrophys. J.} {\bf 544}, 937-943. 
\bibitem[Wetherill(1980)]{wetherill80}
Wetherill, G. W. 1980. 
Formation of the terrestrial planets.
{\it Annu. Rev. Astron. Astrophys.} {\bf 18}, 77-113. 
\bibitem[Whitmire {\it et al.}(1998)]{whitmire}
Whitmire, D. P., J. J. Matese, L. Criswell, and S. Mikkola 1998.
Habitable Planet Formation in Binary Star Systems.
{\it Icarus}, {\bf 132}, 196-203.
\bibitem[Yabushita {\it et al.}(1982)]{yabu82}
Yabushita, S., I. Hasegawa, and  K. Kobayashi 1982. 
The stellar perturbations of orbital elements of long-period comets. 
{\it Mon. Not. R. Astron. Soc.} {\bf 200}, 661-671.  
\end{thebibliography}
\end{document}